\documentclass[a4paper,12pt,nofootinbib,superscriptaddress,showpacs]{revtex4}
\usepackage[frenchb,english]{babel}
\usepackage[applemac]{inputenc}
\usepackage{enumerate}
\usepackage{amsmath}
\usepackage{amsthm}
\usepackage{amssymb}
\usepackage{graphicx}
\usepackage{fancybox}
\usepackage{color}
\usepackage{hyperref}

\newcommand{\be}{\begin{equation}}
\newcommand{\ee}{\end{equation}}
\newcommand{\bal}{\begin{equation} \begin{aligned}}
\newcommand{\eal}{\end{aligned} \end{equation}}
\newcommand{\bi}{\begin{itemize}}
\newcommand{\ei}{\end{itemize}}
\newcommand{\ben}{\begin{enumerate}}
\newcommand{\een}{\end{enumerate}}
\newcommand{\bmat}{\begin{pmatrix}}
\newcommand{\emat}{\end{pmatrix}}

\begin{document}
\baselineskip 17pt

\selectlanguage{english}

\title{Hawking effect in BECs acoustic white holes}

\author{Roberto Balbinot}
\email{balbinot@bo.infn.it}
\affiliation{Dipartimento di Fisica dell'Universit\`a di Bologna and INFN sezione di Bologna, Via Irnerio 46, 40126 Bologna, Italy}
\author{Alessandro~Fabbri}
\email{afabbri@ific.uv.es}
\affiliation{Museo Storico della Fisica e Centro Studi e Ricerche 'Enrico Fermi', Piazza del 
Viminale 1, 00184 Roma, Italy; Dipartimento di Fisica dell'Universit\`a di Bologna, Via Irnerio 46, 40126 Bologna, Italy; Departamento de F\'isica Te\'orica and IFIC, Universidad de Valencia-CSIC, C. Dr. Moliner 50, 46100 Burjassot, Spain}
\author{Carlos Mayoral}
\email{carlos.mayoral@uv.es}
\affiliation{Departamento de F\'isica Te\'orica and IFIC, Universidad de Valencia-CSIC, C. Dr. Moliner 50, 46100 Burjassot, Spain}


\date{\today}

\begin{abstract}
Bogoliubov pseudoparticle creation in a BEC undergoing a WH like flow is investigated analytically in the case of a one dimensional geometry with stepwise homogeneous regions. Comparison of the results with those corresponding to a BH flow is performed. The implications for the analogous gravitational problem is discussed. 
\end{abstract}



\maketitle

\newpage

\section*{Introduction}

Hawking radiation is one of the most spectacular and unexpected predictions of modern theoretical physics.
In 1974 Hawking showed  \cite{hawking}, by combining General Relativity and Quantum Mechanics, that stationary (even static) black holes are expected to emit thermal radiation at a temperature proportional to the surface gravity
of their horizon. 
This had a profound impact on theoretical physics leading to a beautiful synthesis between gravity and 
thermodynamics, cathalized by quantum mechanics.   
Unfortunately, the astrophysical relevance of Hawking radiation is negligible: solar mass black holes should 
radiate at a temperature of the order of $10^{-6}\ K$, much below the cosmic microwave background.
There is no hope to identify such a tiny signal in the sky. Nevertheless there is no doubt that Hawking's 
result remains a milestone in the quest of unifying gravity and quantum mechanics.

The basic feature of Hawking radiation is that it is the result of a pairs production process which proceeds
by the conversion of vacuum fluctuations into on shell particles.
The possibility of having this pairs production in a stationary or even static setting lies in the existence 
of negative frequency modes. These indeed exist inside a BH horizon. They are trapped modes which are
associated with the so called ``partners'' of the Hawking thermal quanta, that reach the asymptotic region 
far from the black hole \cite{mapar}.

In Fig. 1 there is a sketch of the light cones in a black hole (BH) spacetime and in Fig. 2 of the modes 
involved in the Hawking process. These are outgoing, i.e. in principle right-moving. However inside the BH 
horizon they are unable to escape and are forced by the strong gravitational field to move to the left towards the internal singularity.

\begin{figure}
\centering \includegraphics[angle=0, height=1.5in] {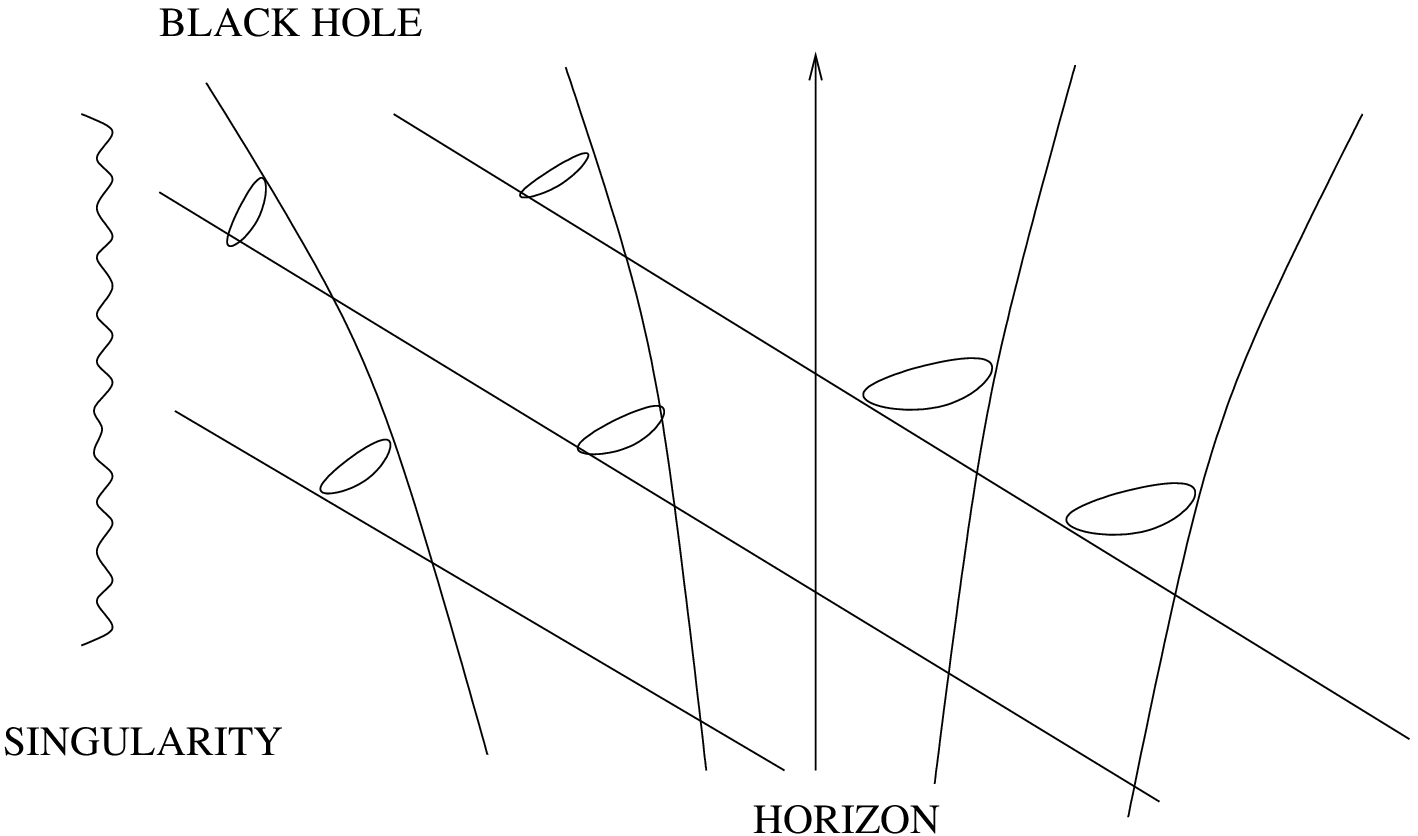}
\caption{Light-cones in a black hole spacetime }
\centering \includegraphics[angle=0, height=1.5in] {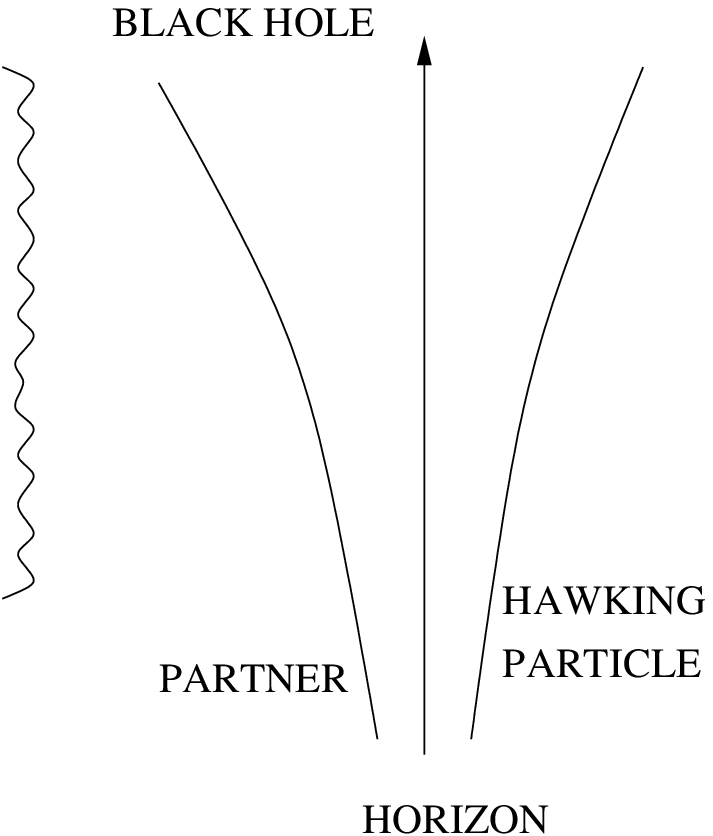}
\caption{Modes involved in the Hawking process}
\end{figure}


Note that the singularity plays no role at all in the Hawking process. It is just the future end-point of any
physical trajectory inside the BH: it has no causal effect on the spacetime.
One should also mention that the ingoing (left moving) modes are basically unaffected by the presence of 
the horizon, and participate only marginally to the process through backscattering.

Finally, Hawking radiation does not depend on the details of the formation of the BH (stellar collapse for 
instance). These determine only an initial transient emission which rapidly (typically in an exponential way) 
decays in time, leaving a steady (ignoring backreaction effects) thermal radiation.
This is Hawking radiation, which depends only (as required by the ``no-hair'' theorem) on mass, charge and 
angular momentum of the BH through the surface gravity \cite{hawking}.

The most serious drawback in Hawking's result is the so called ``transplanckian'' problem \cite{jacobson}. Because of the 
exponential redshift suffered by particles modes constituting Hawking radiation in their journey from
the horizon region to infinity, only very high energy modes (whose frequency is much larger that the Planck one)
created near the horizon manage to reach the asymptotic region.
This fact makes people rather uneasy since it requires an extrapolation of our knowledge of physics 
to an unkown land characterized by scales less than $10^{-33}\ cm$, where quantum gravity effects might in 
principle completely alter the picture. We shall come back to this point later.

We move now to white holes (WH). These are a kind of time reversal of BHs. In Fig. 3 the propagation of
light (light cones) in a WH is depicted. One sees that the horizon once again acts as a semipermeable membrane. 
For a BH (Fig. 1) it is permeable only from outside to inside. For a WH (Fig. 3) the horizon is permeable only
from inside to outside.

\begin{figure}
\centering \includegraphics[angle=0, height=1.5in] {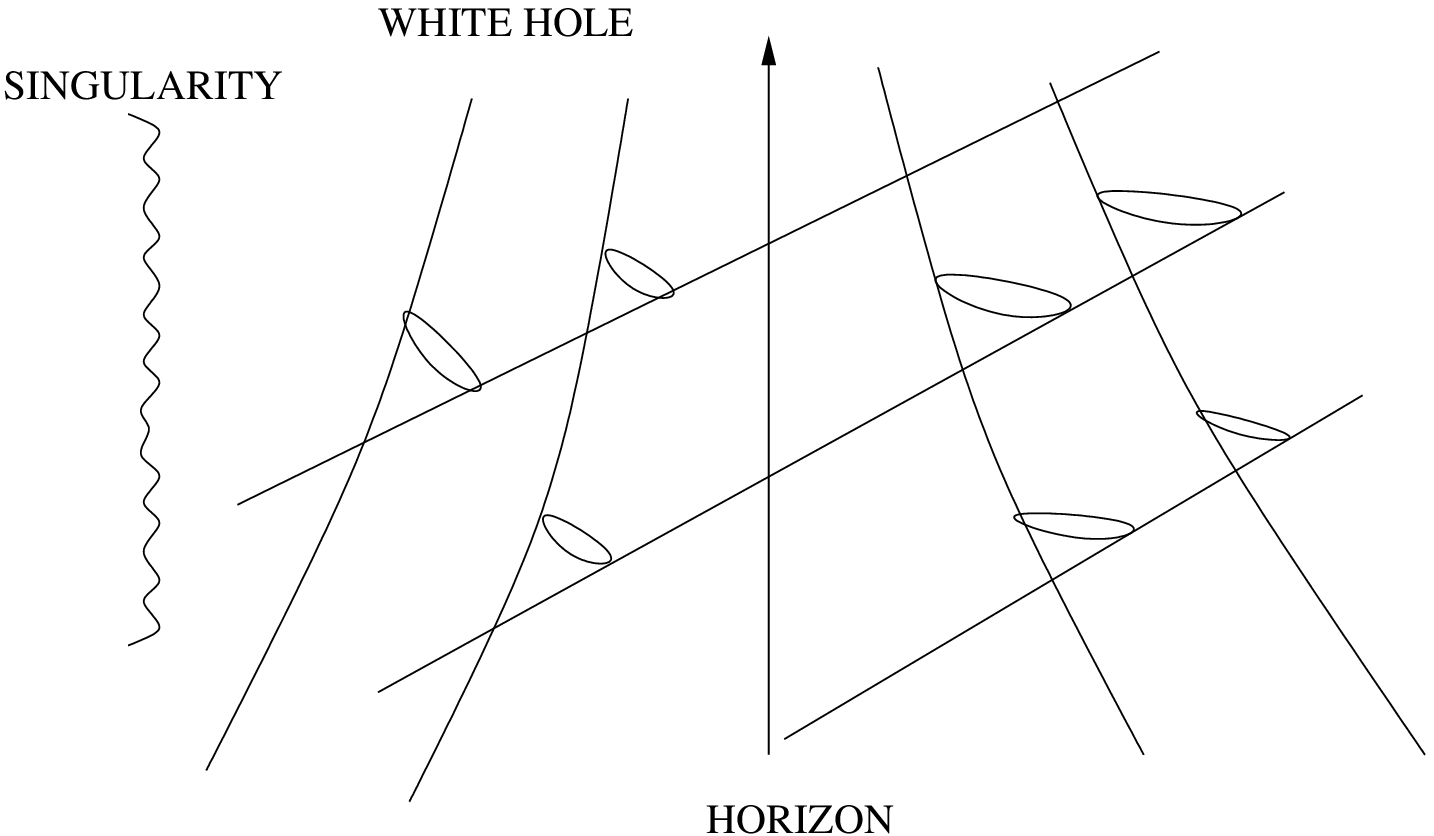}
\caption{Light-cones in a white hole spacetime }
\centering \includegraphics[angle=0, height=1.5in] {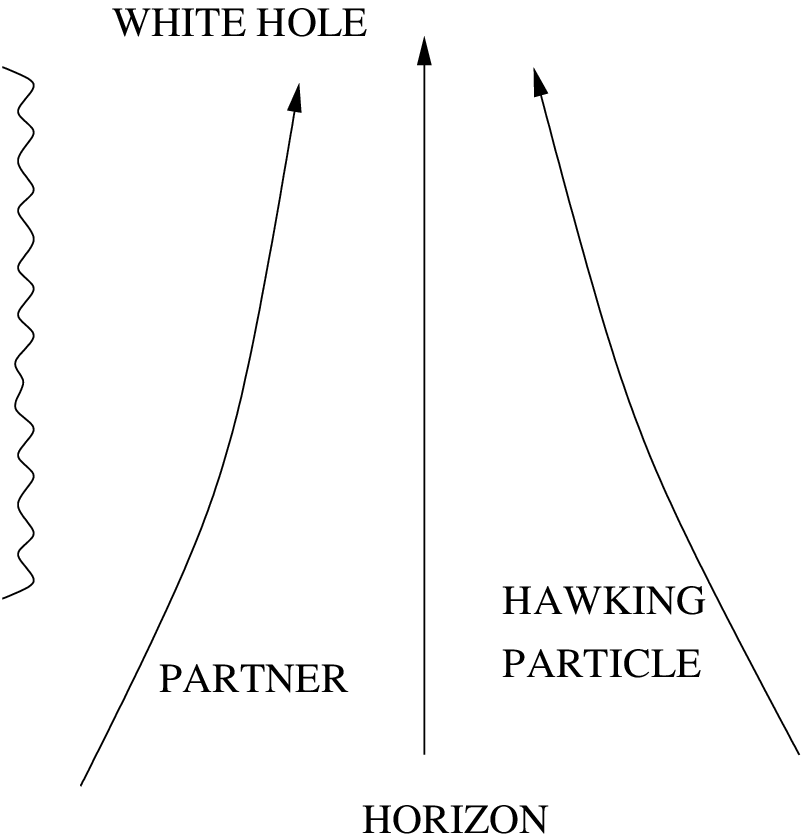}
\caption{Hawking process in a white hole spacetime}
\end{figure}

While a BH swallows everything which falls inside the horizon, a WH expels everything that lies inside the horizon, 
the ultimate origin of this matter being the singularity. 
Note that nothing can penetrate the WH horizon from outside. 
Ingoing physical trajectories pile up along the horizon which is an infinite, this time, blueshift null surface.
This causes WHs to be unstable and therefore of limited interest in gravity as compared to BHs
\cite{eardley}.

Nevertheless one can infer that, beacuse inside the WH horizon there are negative energy modes, there should exist a
sort of Hawking radiation also for them. The modes now involved are the ingoing (potentially left moving) 
ones, as can be seen in Fig. 4. Here we see that the transplanckian problem appears in the future since the 
relevant modes pile up along the horizon and become infinitely blueshifted.
Unlike the BH case, the process seems to be highly sensible to initial conditions.

A much more serious problem is the presence of the singularity which now lies in the past and causally affects
the rest of the spacetime, making every physical prediction about WHs quite arbitrary. 

As we have already stressed, at the heart of the Hawking effect is the existence of negative energy states
which allows pair creation out of the vacuum even in a stationary setting. These peculiar states exist
inside the horizons of both BHs and WHs. But this is not at all an exclusive feature of these exotic systems.
It can be found in many other ``more normal'' systems of different nature, most notable in fluids whose
motion becomes supersonic in some region \cite{unruh}.

One can envisage two flows configurations which are the fluid analogues of the gravitational BH and WH, see Fig. 5.
It is easily realised that sound waves propagation in the two fluid configurations proposed has the same
qualitative behaviour as light propagation in a BH and a WH respectively. One indeed sees that in Fig. 5a
sound waves in the supersonic region 
are trapped and dragged to the left by the flow as it happens
for light inside a BH (see Fig. 1). Fig. 5a represents a sonic BH and similarly Fig. 5b represents a sonic WH, in which sound waves in the supersonic region 
are forced to move towards
the subsonic 
one.
The place where the flows turn from subsonic to supersonic (or viceversa) plays the role of the horizon 
(sonic horizon).
The existence, here, of a trapped region bounded by the horizon, unlike the gravitational case (singularity
theorems \cite{Hawking:1973uf}), does not imply the presence of a singularity, and this makes in particular the study of sonic WHs 
physically  meaningful.

\begin{figure}[htbp]
\begin{center}
\resizebox{!}{4cm}{\includegraphics{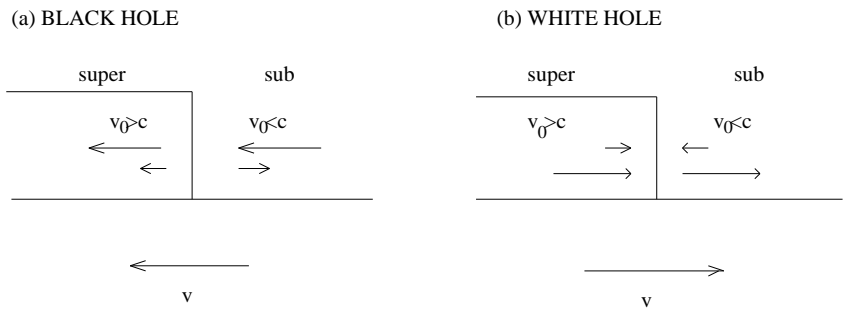}}
\caption{Flow configurations analogue to a BH (a) and a WH (b).}
\label{fig:BH_WH_fluid}
\end{center}
\end{figure}
 
Because there are negative energy states inside sonic horizons, one expects the presence of an analogue Hawking 
radiation (of phonons) in both sonic BH and WH. 

The ``transplanckian problem'' associated to Hawking radiation translates in the fluid context by the fact that 
the relevant sound modes have characteristic wavelengths smaller than the fluid constituents distance,    
which clearly makes no sense at all. 
While there is no hope to address properly the transplanckian problem in gravity (we do not have yet a consistent 
quantum gravity theory), in the fluid case one often has a complete and well tested microscopic quantum
mechanical description of the system at the molecular and atomic scale.

The very existence and related features of the analogue Hawking radiation can then be investigated at a
fundamental level and the theoretical predictions hopefully tested in laboratory experiments.
The most favourable experimental setting to this end is probably offered by Bose-Einstein condensates  \cite{pistri},
where the generally huge difference between the intrinsic temperature and the Hawking one can be 
significantly reduced. 

In BECs the dispersion relation ceases to be linear at high waves momenta and becomes ``superluminal''.
The behaviour of the relevant modes involved in Hawking radiation in a BEC significantly differs from the
one corresponding to the gravitational case and depicted in Fig. 2 (BH) and Fig. 4 (WH).
Indeed for the BEC BHs and WHs configurations one finds, respectively, the behaviours in Fig. 6a and Fig. 6b.

\begin{figure}[htbp]
\begin{center}
\resizebox{!}{5cm}{\includegraphics{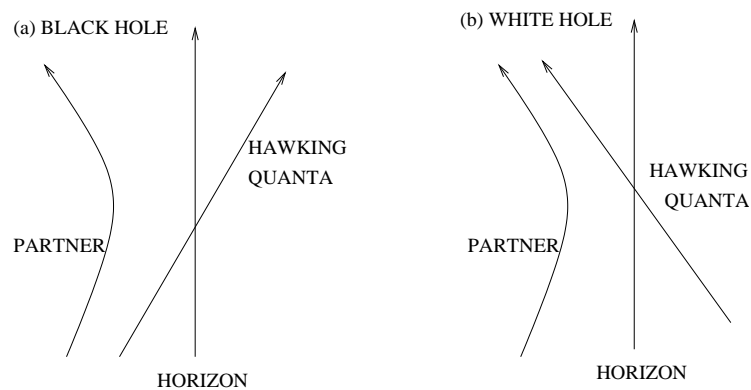}}
\caption{Behavior of the Hawking quanta and the partner in a BH (a) and a WH (b) in the case of a supersonic dispersion relation.}
\label{fig:BH_WH_hawking disp}
\end{center}
\end{figure}

One clearly sees that there is no longer an infinite blueshift (in the past for BH, in the future for WH),
the modes no longer pile up along the horizon and no exponentially large frequencies are involved.
The ``transplanckian problems'' simply do not exist for BEC acoustic BH/WHs, the modified dispersion relation
has completely removed it.

The aim of this paper is to provide a detailed study of Hawking-like radiation in BEC undergoing a WH like supersonic flow \cite{wh}. To allow an analytical (rather than numerical) treatment, which leads to a clear insight in the physical process responsible for the emission, an idealised setting is considered in which a step-like discontinuity in the sound velocity separates a subsonic homogeneous region from a supersonic one \cite{step, Balbinot:2012xw}. Similarity and differences between BH and WH emission will be outlined. 

\section*{Stationary stepwise homogeneous condensates}
\label{Sett}

\subsection{General framework}

The theoretical setup we shall use is the same adopted in ref ~\cite{Balbinot:2012xw} for discussing  
Bogoliubov phonon creation ``\`a la Hawking" in a BEC undergoing a BH like flow. We shall outline here
the basic ingredients. \par \noindent
A system of interacting bosons, confined by an external potential $V_{ext}$,  is described in a second quantized formalism by a field operator $\hat \Psi (t,\vec x)$ which annihilates an atom at $t,\ \vec x$.
$\hat\Psi$ satisfies the usual bosonic commutator rules
\begin{equation}\label{etc}
[\hat \Psi (t,\vec x), \hat \Psi^{\dagger}(t,\vec x')]=\delta^3(\vec x- \vec x')\ .
\end{equation}
 The time evolution of $\hat\Psi$ is given in the diluite gas approximation by
\begin{equation}\label{eqex}
i\hbar\partial_t \hat\Psi = \left(-\frac{\hbar^2}{2m}\vec \nabla^2 + V_{ext} + g \hat\Psi^{\dagger}\hat\Psi \right)\hat\Psi\ ,
\end{equation}
 where $m$ is the mass of the atoms and $g$ is the nonlinear atom-atom interaction constant. \par\noindent
At low temperatures ($T\sim 100 nK$), most of the atoms condense in a common ground state which is described by a $c$ number wave function $\Psi_0$ that satisfies Gross-Pitaevski equation
\begin{equation}\label{gp}
i\hbar\partial_t \Psi_0 = \left(-\frac{\hbar^2}{2m}\vec \nabla^2 + V_{ext} + g n \right)\Psi_0\ ,
\end{equation}
where $n=|\Psi_0|^2$ is the number density of the condensate. Linear perturbation around this classical
macroscopic condensate can be parametrized by an operator $\hat\phi$ as follows
\begin{equation}
\hat\Psi=\Psi_0(1+\hat \phi)\ .
\end{equation}
$\hat\phi$ satisfies the Bogoliubov-de Gennes equation
\begin{equation}\label{bdg}
i\hbar  \partial_t   \hat \phi= - \left( \frac{\hbar^2}{2m}\vec \nabla^2 + \frac{\hbar^2}{m}\frac{\vec \nabla \Psi_0 }{\Psi_0} \vec \nabla\right)\hat\phi +mc^2 (\hat\phi + \hat\phi^{\dagger})\ ,
\end{equation}
where we have introduced $c=\sqrt{\frac{gn}{m}}$, the speed of sound, a fundamental quantity in the following discussion. \par \noindent
The configuration we shall consider is very idealized, consisting of two semiinfinite one dimensional stationary homogeneous condensates attached along a step-like discontinuity. In one region, say the left ($x<0$) the flow is supersonic, while in the right one ($x>0$) the flow is subsonic. The discontinuity is located at $x=0$. See Fig. 7. 

\begin{figure}
\centering \includegraphics[angle=0, height=1.5in] {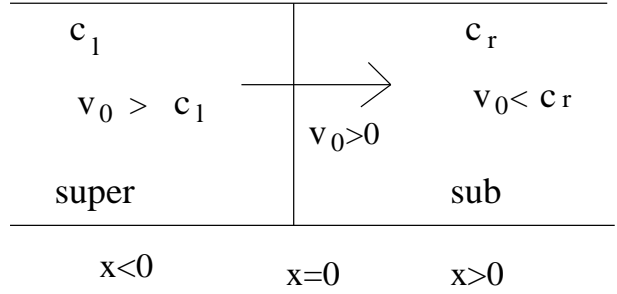}
\caption{Acoustic white hole configuration made of two semiinfinite homogeneous condensates: one supersonic ($x<0$) and one subsonic ($x>0$).
}
\end{figure}

The condensate is supposed to flow from left to right at an everywhere constant velocity $v_0$. 
This configuration, as mentioned in the Introduction, resembles a white hole, since sound waves (i.e. hydrodynamical excitations of the BEC) in the supersonic region are not able to propagate upstream, they are dragged by the flow towards the horizon ($x=0$) and ejected in the subsonic region. Moreover, sound waves in the subsonic region are unable to penetrate the horizon which acts for them as a repulsive surface.
\par \noindent
Furthermore we assume that the condensate has everywhere (i.e. in both regions) a spatial uniform density $n$. One can however have different (constant) sound velocities in the two regions (say, $c_l$ and $c_r$) by changing accordingly the interaction coupling $g$ (remember that  $c=\sqrt{\frac{gn}{m}}$) using the 
so called Feshbach resonance method \cite{pistri}. 
If one changes the external potential from one region to the other as 
\begin{equation}\label{cond}
V_{ext}^l+g^l n=const.=V_{ext}^r+g^r n\ ,
\end{equation}
where $l(r)$ refers to the left (right) region, one has that a simple plane-wave
\be \Psi_0(x,t)=\sqrt{n}e^{ik_0x-iw_0t} \ee
with $\frac{\hbar k_0}{m}=v_0$ and $w_0=\frac{\hbar k_0^2}{2m}+(gn+V_{ext})$ is for all time a solution of the Gross-Pitaveski equation (\ref{gp}) for every $x$. \par \noindent
Concerning the (non hermitean) fluctuation operator $\hat\phi$, owing to the stationarity of our configuration, it can be decomposed in a positive and negative frequency part, namely
\begin{equation}\label{frep}
\hat\phi (t,x) =\sum_j \left[ \hat a_j \phi_j (t,x) + \hat a_j^{\dagger} \varphi_j^*(t,x) \right]\ .\end{equation}
$\hat a_i$ and $\hat a_i^\dagger$ are phonons' annihilation and creation operators respectively. They satisfy bosonic commutation relations
$[\hat a_i,\hat a_j^{\dagger}]=\delta_{ij}$. \par\noindent
Inserting the decomposition (\ref{frep}) into the Bogoliubov-de Gennes equations (\ref{bdg}) and its hermitean conjugate, we see that the modes satisfy the coupled equations
\begin{eqnarray}\nonumber
\left[ i(\partial_t + v_0\partial_x) + \frac{\xi c}{2} \partial_x^2 -\frac{c}{\xi} \right] \phi_j &=& \frac{c}{\xi}
\varphi_j\ , \\
\left[ -i (\partial_t + v_0\partial_x) + \frac{\xi c}{2}\partial_x^2 - \frac{c}{\xi}\right] \varphi_j &=&\frac{c}{\xi} \phi_j  \ , \label{cde}
\end{eqnarray}
where $\xi\equiv \hbar/mc$ is the so called healing length, which, as we shall see, sets the fundamental scale in our system, it is the analogous of the Planck length in gravity. \par \noindent
As said above, because of the stationarity of the flow, the time dependence for both $\phi_j$ and
$\varphi_j$ can be taken as $e^{-iwt}$, where $w$ is the conserved frequency of the modes as measured by a static (laboratory) observer. 
So $w$ will label our modes, i.e. we shall work with modes at fixed $w$. 
The modes are normalized according to 
\begin{equation}
\label{nor}
\int dx [\phi_w\phi_{w'}^* - \varphi_w^*\varphi_{w'}]=\pm\frac{\delta_{ww'}}{\hbar n}\ ,\end{equation}
where the $+(-)$ sign refers to positive (negative) norm states. \par \noindent
Piecewise homogeneity allows in each region, $r$ or $l$, the solutions for the modes to be given in the form of plane waves 
\begin{eqnarray}
\phi_w &=& D(w) e^{-iwt+ikx}\ ,\\
\varphi_w &=& E(w) e^{-iwt+ikx} \ ,
\end{eqnarray}
where $D(w)$ and $E(w)$ are normalization factors to be determined using eq. (\ref{nor}). 
Inserting the plane wave solutions in the Bogoliubov- de Gennes equations (\ref{cde})  we obtain
\begin{eqnarray}\nonumber
\left[ (w-v_0k) - \frac{\xi c k^2}{2}  -\frac{c}{\xi} \right] D(\omega) &=& \frac{c}{\xi} E(\omega)\ , \\
\left[ - (w-v_0k) - \frac{\xi c k^2}{2} - \frac{c}{\xi}\right] E(\omega) &=& \frac{c}{\xi} D(\omega)  \ .\label{gupa}
\end{eqnarray}
This homogeneous linear system has nontrivial solution if the associated determinant vanishes. This gives the dispersion relation
\begin{equation}\label{nrela}
(w-v_0k)^2=c^2\left(k^2+ \frac{\xi^2 k^4}{4}\right),
\end{equation}
which can also be rewritten as 
\begin{equation}\label{relk}
w-v_0k= \Omega_{\pm}(k)=\pm c\sqrt{k^2+\frac{\xi^2k^4}{4}}\ .
\end{equation}
$\Omega_{\pm}(k)$ is the (nonconserved) frequency of the modes as measured in a frame comoving with the fluid. The $\pm$ sign in eq. (\ref{relk}) labels the positive ($+$) and negative ($-$) branches of the dispersion relation. The normalization condition (\ref{nor}) gives 
\begin{equation}\label{nodia}
 |D(\omega)|^2 - |E(\omega)|^2=\pm{1\over 2\pi \hbar n}\Big|\frac{dk}{dw}\Big|\
\end{equation}
which, with the help of eqs. (\ref{gupa}), fixes the normalization coefficients as 
\begin{eqnarray}\label{eq:normdispersion}
D(\omega) &=&  \frac{\omega -v_0 k+\frac{c\xi k^2}{2}}{\sqrt{4\pi \hbar n c\xi k^2\left| (\omega-v_0k) \left(\frac{dk}{d\omega}\right)^{-1} \right| }},\nonumber\\
E(\omega) &=& -\frac{\omega -v_0 k-\frac{c\xi k^2}{2}}{\sqrt{4\pi \hbar nc\xi k^2\left| (\omega-v_0k) \left(\frac{dk}{d\omega}\right)^{-1} \right| }}.
\end{eqnarray}
Using these explicit forms of the normalization coefficients, one realizes that positive norm states belong to the positive branch of the dispersion relation, negative norm states to the negative branch.
\par\noindent
From eq. (\ref{relk}) one can also see that to a $w>0$ negative norm solution of eq. (\ref{cde}), one can 
associate a positive norm solution with $w<0$ (and $k\to -k$). One can then work with only with $w>0$ modes, identifying $w>0$ negative norm states (if present) with negative frequency excitations. \par\noindent
In the most familiar situation of a subsonic flow, for $w>0$ one has only positive norm states (negative norm
states have $w<0$). However, when the flow is supersonic we will see that for a given $w>0$ there can be, in addition to positive norm modes, also negative norm ones which are physically interpreted as described above. \par \noindent
Coming back to the dispersion relation, eq. (\ref{relk}), we see that for small $k$ (i.e. $k\ll 1/\xi$) the dispersion relation is linear $w-v_0k=\pm ck$. This is the hydrodynamical regime ($\xi\to 0$) with solutions
$k_u=\frac{w}{v_0+c},\ k_v=\frac{w}{v_0-c}$ leading to $v_g\equiv \frac{dw}{dk}=v_0\pm c$ for the group velocity of the two modes. In general, at fixed $w$, eq. (\ref{relk}) is fourth order in $k$, which admits four roots $k_w^{(i)}$ ($i=1,2,3,4$). So we can write the general solution of the Bogoliubov-de Gennes eq. (\ref{cde}) in each region as a linear combination of four plane waves with the above wave vector $k_w^{(i)}$, i.e. \begin{eqnarray}
\phi_w&=&e^{-iwt}\sum_{i=1}^{4}A_i^{(w)}D_i(w)e^{ik_w^{(i)}x}\ , \nonumber \\
\varphi_w&=&e^{-iwt}\sum_{i=1}^{4}A_i^{(w)}E_i(w)e^{ik_w^{(i)}x}\ , \label{soli} 
\end{eqnarray}
where the $A_i(w)$ are the amplitudes of the modes, not to be confused with the normalization coefficients
$D_i(w)$ and $E_i(w)$ which are fixed by the normalization condition and are given by eqs. (\ref{eq:normdispersion}) with $k=k_w^{(i)}$. 
Note that the amplitudes $A_i(w)$  are the same for both $\phi_w$ and $\varphi_w$ as required by the equations of motion (\ref{cde}).\par \noindent
Once the roots of the dispersion relation are found in each region, one can construct the corresponding solution, eqs. (\ref{soli}), in the left and right region with the corresponding left and right amplitudes 
$A_i^l(w),\ A_i^r(w)$. 
Note that the explicit form and features of the four roots will depend on the sub or supersonic character of the flow and so will be significantly different in the two regions. \par \noindent
Now, the equations of motion, eqs. (\ref{cde}) require $\phi$ and $\varphi$ and their space derivatives to be continuous across the step discontinuity at $x=0$, namely 
\begin{equation}\label{matchingaa}
[\phi]=0,\, [\phi']=0,\, [\varphi]=0,\, [\varphi']=0,
\end{equation}
where $[f(x)]=\lim_{\epsilon\to 0} [f(x+\epsilon)-f(x-\epsilon)]$ and a prime means $\frac{d}{dx}$.
The four matching conditions, eqs. (\ref{matchingaa}), establish a linear relation between the four left
amplitudes $A_i^l$ and the four right amplitudes $A_i^r$
\begin{equation}
A_i^l=M_{ij}A_j^r
\end{equation}
where $M_{ij}$ is the $4\times 4$ ``Matching Matrix", not to be confused with the scattering matrix to be introduced later. 

\subsection{Subsonic region}

Let us first find the roots of the dispersion relation (\ref{relk}) in the subsonic (right) region. This is graphically displayed in Fig. 8, where the solid line corresponds to the positive branch, the dashed line to the negative branch.

\begin{figure}
\centering \includegraphics[angle=0, height=2.0in] {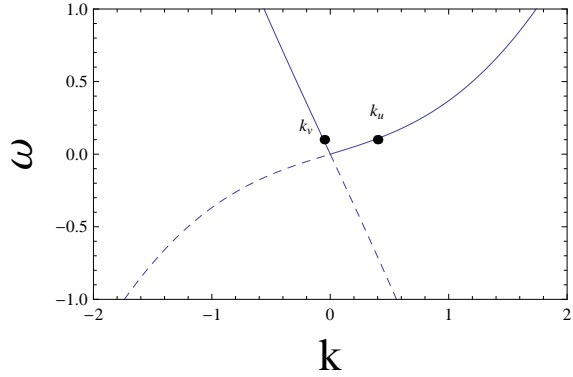}
\caption{Dispersion relation in the subsonic (right) region.
}
\end{figure}

So, for $w>0$ two real solutions exist, both belong to the positive norm branch, namely $k_u^r$ with positive group velocity (propagation towards the right, along the flow) and $k_v^r$ with negative group velocity (propagation towards the left, against the flow). These roots admit a perturbative expansion for small $w$ as
\begin{eqnarray}\label{uv}
&&k_v^r=\frac{\omega}{v_0-c_r}\left(1+\frac{c_rw^2\xi_r}{8(v_0-c_r)^3}+O(z_r^4)\right)\ ,\nonumber\\
&&k_u=\frac{\omega}{v_0+c_r}\left(1-\frac{c_rw^2\xi_r}{8(v_0+c_r)^3}+O(z_r^4)\right)\ ,
\end{eqnarray}
where the dimensionless expansion parameter used is $z_r\equiv {\xi_r \omega\over c_r}$. The order $0$ terms are just the hydrodynamical modes discussed before. 
The other two solutions of the dispersion relation are a couple of complex conjugate roots $k_{\pm}^r$.
The one with positive imaginary part ($k_+^r$) represents a decaying mode as $x\to +\infty$. The other, $k_-^r$, with negative imaginary part, a growing mode. 
These two complex roots do not exist in the hydrodynamical approximation. Their presence is a consequence of the nonlinearity of the dispersion relation which at $k\gg\xi$ becomes quadratic (single particle excitation). $k_{\pm}^r$ have a small $w$ expansion
\begin{equation}\label{decaying}
k_{\pm}^r=\frac{\omega v_0}{c_r^2-v_0^2}\left[1-\frac{(c_r^2+v_0^2)c_r^2w^2\xi_r^2}{4(c_r^2-v_0^2)^3}+O(z_r^4)\right]\pm \frac{2i\sqrt{c_r^2-v_0^2}}{c_r\xi_r}\left[1+\frac{(c_r^2+2v_0^2)c_r^2w^2\xi_r^2}{8(c_r^2-v_0^2)^3}+O(z_r^4)\right]\ .
\end{equation}
Having the four roots, we can write down the solution for the modes equations (\ref{cde}), in the subsonic (right) region as  
\begin{equation}\label{decuno}
\phi_{\omega}^{r} = e^{-i\omega t}\left[A_v^{r}D_v^{r}e^{ik_v^{r}x}+A_u^{r}D_u^{r}e^{ik_u^{r}x}
+A_+^{r}D_{+}^{r}e^{ik_{+}^{r}x}+A_-^{r}D_{-}^{r}e^{ik_{-}^{r}x}\right]\ ,
  \end{equation}
and similarly for $\varphi_w^r$ (replace $D_i^r\to E_i^r$). 

\subsection{Supersonic region}

Let's move to the supersonic ($x<0$, left) region. The dispersion relation (see Fig. 9) shows a completely different pattern, as compared to the subsonic case.

\begin{figure}
\centering \includegraphics[angle=0, height=2.0in] {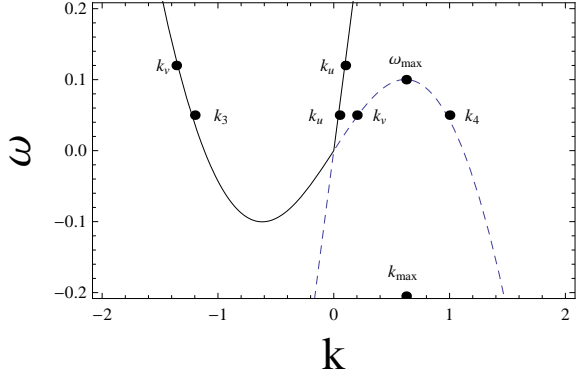}
\caption{Dispersion relation in the supersonic (left) region.
}
\end{figure}

One sees that there exists a maximum frequency $w_{max}$ above which the description mimics exactly what happens in the subsonic case: two oscillatory modes $k_u^l$ and $k_v^l$ belonging both to the positive
branch and propagating to the right and to the left respectively and two complex conjugate roots $k_{\pm}^l$. Note however that, although we are in the supersonic region, i.e. in the white hole, the $k_v^l$ mode is able to propagate to the left, i.e. upstream, notwithstanding the BEC is flowing supersonically 
in the right direction. The group velocity of this mode is clearly bigger than $v_0$ and also see from Fig. 9 
that the corresponding root lives well outside the linear (hydrodynamical) regime. $w_{max}$ is the value of $w$ for which $\frac{dw}{dk}|_{k_{max}}=0$ where   
\begin{equation}
k_{max}=-\frac{1}{\xi_{l}}\left[-2+\frac{v_0^2}{2c_l^2}+\frac{v_0}{2c_l}\sqrt{8+\frac{v_0^2}{c_l^2}}\right]^{1/2}\ .
\end{equation}
So for $w>w_{max}$ the solution of the mode equations (\ref{cde}) can be given in a way completely analogous to what was done in the subsonic case, namely eq. (\ref{decuno}). 
One can then impose the jump conditions (\ref{matchingaa}) on the left and right solutions across the discontinuity.
We shall not discuss this case further since it just leads to a simple scattering process for the mode as would happen in a sub-sub configuration. \par\noindent
From now on we shall restrict our discussion on modes with $w<w_{max}$ which will take us to a much richer  physical outcome. 
Looking at Fig. 9 one sees that there are now four real roots $k_u^l$, $k_v^l$, $k_3^l$, $k_4^l$ each corresponding to an oscillatory mode. The first two are the familiar hydrodynamical-like modes
\begin{eqnarray}
&&k_v^l=\frac{\omega}{v_0-c_l}\left[1+\frac{c_lw^2\xi_l^2}{8(v_0-c_l)^3}+O(z_l^4)\right]\ ,\nonumber\\
&&k_u^l=\frac{\omega}{v_0+c_l}\left[1-\frac{c_lw^2\xi_l^2}{8(v_0+c_l)^3}+O(z_l^4)\right]\ ,
\end{eqnarray}
where $z_l\equiv \frac{\xi_l w}{c_l}$.
But while the $k_u^l$ mode has, as before, positive group velocity and belongs to the positive norm branch, the $k_v^l$ one, unlike what happens in the subsonic case, has positive group velocity and furthermore it belongs to the negative norm branch. The corresponding excitations are negative energy ($w<0$) phonons dragged by the fluid to the right, towards the horizon, they are no longer able to propagate against the flow.
The other two roots $k_3^l$, $k_4^l$ are not hydrodynamical modes
\begin{equation}
k_{3(4)}=-\frac{\omega v_0}{v_0^2-c_l^2}\left[1+\frac{(c_l^2+v_0^2)c_l^2w^2\xi_l^2}{4(v_0^2-c_l^2)^3}+O(z_l^4)\right]
-(+)\frac{2\sqrt{v_0^2-c_l^2}}{c_l\xi_l}\left[1-\frac{(c_l^2+2v_0^2)c_l^2w^2\xi_l^2}{8(v_0^2-c_l^2)^3}+O(z_l^4)\right]\ .
\end{equation}
These two additional oscillatory roots are the analytical continuation of the decaying and growing complex roots ($k_{\pm}$) present in the subsonic regime (see eqs. (\ref{decaying})). 
These two modes have both negative group velocity, i.e. notwithstanding the supersonic character of the flow, they are both able to propagate against the flow (i.e. to the left). Clearly their group velocity is bigger than $v_0$ ($|v_g|>v_0$). 
Finally note that $k_3^l$ belongs to the positive norm branch while $k_4^l$ to the negative one.
Given this, the general solution of the modes equation in the left (supersonic) region can be given as (for $w<w_{max}$) 
\begin{equation}
  \phi_{\omega}^{l} = e^{-i\omega t}\left[A_v^{l}D_v^{l}e^{ik_v^{l}x}+A_u^{l}D_u^{l}e^{ik_u^{l}x}+A_3^{l}D_{3}^{l}e^{ik_3^{l}x}
+A_4^{l}D_{4}^{l}e^{ik_4^{l}x}\right] \ .
  \end{equation}
A similar expansion holds for $\varphi_w$ with $D^i_j\to E^i_j$, while the amplitudes $A^i_j$ remain the same as stressed before.

\subsection{The matching matrix $M_{ij}$}

The four jump conditions at $x=0$ (eqs. (\ref{matchingaa})) can be written as a linear system 
\begin{equation}
\label{wrwl}
W_l\left(
     \begin{array}{c}
       A_v^l \\
       A_u^l \\
       A_3^l \\
       A_4^l \\
 \end{array}
   \right)=W_r\left(
                \begin{array}{c}
                  A_v^r \\
       A_u^r \\
       A_+^r \\
       A_-^r \\
                \end{array}
              \right),
\end{equation}
 where the $4\times 4$ matrices  $W_{l}$ and $W_{r}$ read respectively  
\begin{equation}
\label{eq:wl}
W_{l}=\left(
     \begin{array}{cccc}
       D_v^{l} & D_u^{l} & D_{3}^{l} & D_{4}^{l}\\
       ik_v^{l}D_v^{l} & ik_u^{l}D_u^{l} & ik_3^{l}D_{3}^{l} & ik_4^{l}D_{4}^{l} \\
       E_v^{l} & E_u^{l} & E_{3}^{l} & E_{4}^{l} \\
       ik_v^{l}E_v^{l} & ik_u^{l}E_u^{l} & ik_3^lE_{3}^{l(r)} & ik_4^{l}E_{4}^{l}  \\
\end{array}\right)\ ,
\end{equation}
\begin{equation}
\label{eq:wl}
W_{r}=\left(
     \begin{array}{cccc}
       D_v^{r} & D_u^{r} & D_{+}^{r} & D_{-}^{r}\\
       ik_v^{r}D_v^{r} & ik_u^{r}D_u^{r} & ik_+^{r}D_{+}^{r} & ik_-^{r}D_{-}^{r} \\
       E_v^{r} & E_u^{r} & E_{+}^{r} & E_{-}^{r} \\
       ik_v^{r}E_v^{r} & ik_u^{r}E_u^{r} & ik_+^rE_{+}^{r} & ik_-^{r}E_{-}^{r}  \\
\end{array}\right)\ . 
\end{equation}
Multiplying both sides of eq. (\ref{wrwl}) by $W_l^{-1}$  one gets
\begin{equation}
     \left( \begin{array}{c}
       A_v^l \\
       A_u^l \\
       A_3^l \\
       A_4^l \\
     \end{array} \right)
   =M \left(
                \begin{array}{c}
                             A_v^r \\
       A_u^r \\
       A_+^r \\
       A_-^r \\
                \end{array}
              \right) \ ,
\end{equation}
where $M =W_l^{-1}W_r$ is the matching matrix. 

\subsection{The ``in" scattering basis}

We now proceed, starting from the modes discussed in the previous sections, to construct a complete and 
orthonormal basis, with respect to the scalar product eq. (\ref{nor}), to be used to expand the field operator
$\hat\phi$ as indicated in eq. (\ref{frep}). This can be done in two ways. Either we choose incoming modes to build up the basis, i.e. modes that in the remote past ($t\to -\infty$) start their journey from left or right spatial infinity ($x=-\infty,\ x=+\infty$) and propagate towards the discontinuity at $x=0$, or we choose outgoing modes that at $t=+\infty$ propagate towards $x=\pm\infty$. \par\noindent
For $w>w_{max}$ we have two incoming modes ($k_u^l,\ k_v^r$) that are scattered (transmitted and reflected) in two outgoing modes ($k_v^l,\ k_u^r$). As said this case is trivial nd we shall not discuss it.
\par \noindent
For $w<w_{max}$ there are three incoming modes ($k_u^l,\ k_v^l,\ k_v^r$) and three outgoing modes ($k_3^l,\ k_4^l,\ k_u^r$). 
\par\noindent
We begin with the construction of the ``in" basis. We define the $\phi_{vr}^{in}(w)$ scattering mode as following: an initial left moving unit amplitude $v$ mode coming at $t=-\infty$ from the right subsonic region (i.e. $D_v^re^{-iwt+ik_v^rx}$) is scattered by the discontinuity at $x=0$ and generates a reflected right moving $k_u^r$ mode with amplitude $A_u^r$ (i.e. $A_u^rD_u^re^{-iwt+ik_u^rx}$) in the right region, and two transmitted $k_3^l$, $k_4^l$ modes, with amplitudes $A_3^l$ and $A_4^l$ respectively, in the left region (i.e.   $A_3^lD_3^le^{-iwt+ik_3^lx}$ and $A_4^lD_4^le^{-iwt+ik_4^lx}$) . 
The construction is not complete since we need to include in the right (subsonic) region the complex decaying mode $k_+^r$ (i.e. $A_+^rD_+^re^{-iwt+ik_+^rx}$). The growing mode is not included as it blows up at $x=+\infty$. The construction of the $\phi_{vr}^{in}(w)$ basis vector is depicted in Fig. 10.

\begin{figure}
\centering \includegraphics[angle=0, height=1.5in] {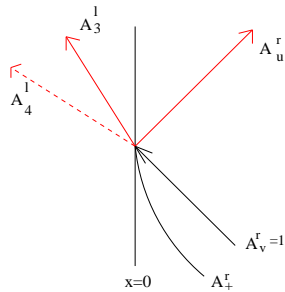}
\caption{Mode $\phi_{v,r}^{in}(w)$.
}
\end{figure}

We stress again that $k_3$ and $k_4$ propagate upstream in the supersonic region and that $k_4$ belongs to the negative norm branch. It represents the so called ``anomalous" transmission and describes, as said, negative energy phonons. 
The associated amplitudes of the modes can be evaluated using the matching matrix $M$ by solving the following system 
\begin{equation}
\label{modevrin}
     \left( \begin{array}{c}
       0 \\
       0  \\
       A_3^l\\
       A_4^l\\
     \end{array} \right)
   = M\left(
                \begin{array}{c}
                  1  \\
                  A_u^r \\
                  A_+^r\\
                  0
                \end{array}
              \right) \ .
\end{equation}
The leading order results for small $w$ read (we have set for simplicity $\hbar=m=1$) 
\begin{eqnarray}
A_u^r &\simeq & \frac{v_0-c_r}{v_0+c_r}\equiv S_{ur,vr}\ , \nonumber \\
A_3^l  &\simeq &\frac{\sqrt{2c_r}(v_0^2-c_l^2)^{3/4}(v_0-c_r)}{\sqrt{w}\sqrt{c_r^2-v_0^2}(c_l^2-c_r^2)}
(\sqrt{c_r^2-v_0^2}+i\sqrt{v_0^2-c_l^2})\equiv S_{3l,vr}\ ,\nonumber \\
A_4^l &\simeq & -A_3^{*l} \Rightarrow S_{4l,vr} \simeq -S_{3l,vr}^* \ .\label{sur}
\end{eqnarray}
The notation introduced is quite simple: the second couple of indices describes the incoming channel ($vr$) while the first couple of indices labels the outgoing channel. We will not need the exlicit form of $A_+^r$. 
The amplitudes of the propagating modes can be shown to satisfy the unitarity relation
\be \label{ure}|A_u^r|^2+|A_3^l|^2-|A_4^l|^2=1\ . \ee
The minus sign in front of the $A_4^l$ term in eq. (\ref{ure}) comes from the fact that the $k_4$ mode has negative norm.\par\noindent
The second basis vector, the scattering mode $\phi_{ul}^{in}(w)$, is constructed following the process depicted in Fig. 11.

\begin{figure}
\centering \includegraphics[angle=0, height=1.5in] {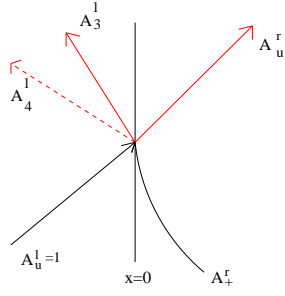}
\caption{Mode $\phi_{ul}^{in}(w)$.
}
\end{figure}

 It corresponds to a unit amplitude positive norm right moving $u$ mode coming from the (left) supersonic region (i.e. $D_u^le^{-iwt+ik_u^lx}$) which is reflected into a positive norm $k_3^l$ mode 
with amplitude $A_3^l$ (i.e. $A_3^lD_3^le^{-iwt+ik_3^lx}$) and a negative norm mode $k_4^l$ with amplitude $A_4^l$ (i.e. $A_4^lD_4^le^{-iwt+ik_4^lx}$). In addition there is a transmitted $k_u^r$ mode in the subsonic region with amplitude $A_u^r$ moving to the right (i.e. $A_u^rD_u^re^{-iwt+ik_u^rx}$) and a decaying mode with amplitude $A_+^r$ (i.e. $A_+^rD_+^re^{-iwt+ik_+^rx}$). The amplitudes for this process are found again using the matching matrix $M$
\begin{equation}
\label{modeulin}
     \left( \begin{array}{c}
       0 \\
       1  \\
       A_3^l\\
       A_4^l\\
     \end{array} \right)
   = M\left(
                \begin{array}{c}
                  0  \\
                  A_u^r \\
                  A_+^r\\
                  0
                \end{array}
              \right) \ .
\end{equation}
The leading order in $w$ results are 
\begin{eqnarray}
A_u^r &\simeq & \sqrt{\frac{c_r}{c_l}}\frac{v_0+c_l}{v_0+c_r}  \equiv S_{ur,ul}\ ,\nonumber \\
A_3^l  &\simeq & \frac{(v_0^2-c_l^2)^{3/4}(c_r-v_0)}{\sqrt{2c_lw}\sqrt{c_r^2-v_0^2}(c_r+c_l)}(\sqrt{c_r^2-v_0^2}+i\sqrt{v_0^2-c_l^2}) \equiv S_{3l,ul}\ ,\nonumber \\
A_4^l &\simeq & -A_3^{l*} \Rightarrow S_{4l,ul} \simeq -S_{3l,ul}^* \ .\label{sul}
\end{eqnarray}
The unitarity relation reads now $|A_u^r|^2+|A_3^l|^2-|A_4^l|^2=1$. \par\noindent
The last scattering mode, completing the ``in" basis, is $\phi_{vl}^{in}(w)$ constructed according to Fig. 12.

\begin{figure}
\centering \includegraphics[angle=0, height=1.5in] {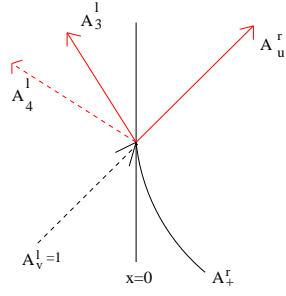}
\caption{Mode $\phi_{vl}^{in}(w)$.
}
\end{figure}

An incoming unit norm, negative norm, $k_v^l$ mode (i.e. $D_v^le^{-iwt+ik_v^lx}$) is reflected in a positive norm $k_3^l$ mode and in a negative norm $k_4^l$ mode with amplitude $A_3^l$ and $A_4^l$ respectively, and transmitted in a $k_u^r$ mode with amplitude $A_u^r$. Furthermore, there is as usual in the subsonic region the decaying mode $k_+^r$ with amplitude $A_+^r$. These amplitudes are found using the matching matrix as follows
\begin{equation}
\label{modevlin}
     \left( \begin{array}{c}
       1 \\
       0  \\
       A_3^l\\
       A_4^l\\
     \end{array} \right)
   = M\left(
                \begin{array}{c}
                  0  \\
                  A_u^r \\
                  A_+^r\\
                  0
                \end{array}
              \right) \ 
\end{equation}
yielding 
\begin{eqnarray}
A_u^r &\simeq & \sqrt{\frac{c_r}{c_l}}\frac{v_0-c_l}{v_0+c_r}  \equiv S_{ur,vl}\ ,\nonumber \\
A_3^l  &\simeq & \frac{(v_0^2-c_l^2)^{3/4}(v_0-c_r)}{\sqrt{2c_lw}\sqrt{c_r^2-v_0^2}(c_l-c_r)}
(\sqrt{c_r^2-v_0^2}-i\sqrt{v_0^2-c_l^2}) \equiv S_{3l,vl}\ ,\nonumber \\
A_4^l &\simeq & -A_3^{l*} \Rightarrow S_{4l,vl} \simeq -S_{3l,vl}^* \ .\label{svl}
\end{eqnarray}
The unitarity condition now reads \be |A_u^r|^2+|A_3^l|^2-|A_4^l|^2=-1\ .\ee
The minus sign on the r.h.s. comes from the fact that the unit amplitude incoming $k_v^l$ mode has negative norm. \par\noindent  
We have now the three ``in" basis vectors ($\phi_{vr}^{in}(w),\ \phi_{ul}^{in}(w),\ \phi_{vl}^{in}(w)$) 
and can expand the field operator $\hat\phi$ as follows
\begin{eqnarray}
\hat\phi&=&\int_{0}^{\omega_{max}}d\omega\Big[\hat a_{vr}^{in}(w)\phi_{vr}^{in}+\hat a_{ul}^{in}(w)\phi_{ul}^{in}(w)+\hat a_{vl}^{in\dagger}(w)\phi_{vl}^{in}(w)
\nonumber \\ &+& \hat a_{vr}^{\dagger in}(w)\varphi_{vr}^{in*}+\hat a_{ul}^{in\dagger}(w)\varphi_{ul}^{in*}(w)+\hat a_{vl}^{in}(w)\varphi_{ul}^{in*}(w) \Big]\ . 
\ \ \ \ \ \ \ \ \ \
\end{eqnarray}
Note in this expansion that the mode $\phi_{vl}^{in}(w)$, being a negative frequency one (negative norm) is associated to a creation operator, while $\phi_{vr}^{in}(w)$ and $\phi_{ul}^{in}(w)$ being positive frequency modes are associated to annihilation operators as usual.\par\noindent
One can alternatively proceed to the construction of the ``out" basis, which consists of unit amplitude
outgoing modes (i.e. modes propagating away from the discontinuity at $x=0$). These three modes
have wave vectors $k_3^l$, $k_4^l$ and $k_u^r$. Starting from each of these one can construct a corresponding basis vector. \par\noindent
We begin defining the scattering mode $\phi_{ur}^{out}(w)$ constructed according to Fig. 13.

\begin{figure}
\centering \includegraphics[angle=0, height=1.5in] {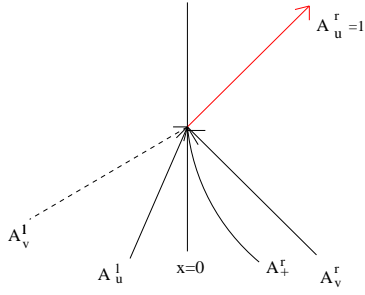}
\caption{Mode $\phi_{ur}^{out}(w)$.
}
\end{figure}
 
This scattering mode is a linear combination of initial $k_u^l,\ k_v^l$ right moving modes coming from the supersonic region and a left moving $k_v^r$ mode coming from the right region with amplitudes $A_u^l,\ A_v^l,\ A_v^r$ respectively, which together with the decaying mode $k_+^r$ (of amplitude $A_+^r$) produce a final right moving $k_u^r$ mode in the subsonic region with unit amplitude.
The corresponding amplitude for the incoming modes can be evaluated using the matching matrix $M$ with a procedure identical to the one used for the ``in" basis. Since we shall not need these amplitudes we shall omit the derivation of their explicit form.\par \noindent
The other two scattering modes, completing the ``out" basis, $\phi_{3l}^{out}(w)$ and $\phi_{4l}^{out}(w)$,
correspond to the process depicted in Fig. 14 and Fig. 15 respectively.

\begin{figure}
\centering \includegraphics[angle=0, height=1.5in] {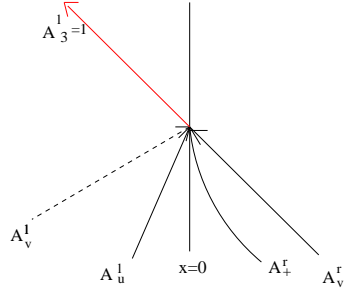}
\caption{Mode $\phi_{3l}^{out}(w)$.
}
\end{figure}

\begin{figure}
\centering \includegraphics[angle=0, height=1.5in] {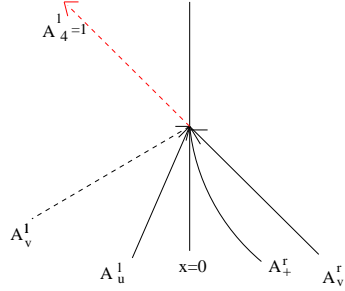}
\caption{Mode $\phi_{4l}^{out}(w)$.
}
\end{figure}

Hence the field operator $\hat\phi$ can be expanded alternatively in the ``out" basis as
\begin{eqnarray}
\hat\phi&=&\int_{0}^{\omega_{max}}d\omega\Big[\hat a_{ur}^{out}(w)\phi_{ur}^{out}+\hat a_{3l}^{out}(w)\phi_{3l}^{out}(w)+\hat a_{4l}^{out\dagger}(w)\phi_{4l}^{out}(w)
\nonumber \\ &+& \hat a_{ur}^{\dagger out}(w)\varphi_{ur}^{out*}+\hat a_{3l}^{out\dagger}(w)\varphi_{3l}^{out*}(w)+\hat a_{4l}^{out}(w)\varphi_{4l}^{out*}(w) \Big]\ . 
\ \ \ \ \ \ \ \ \ \
\end{eqnarray}
As before, note that the negative frequency $\phi_{4l}^{out}(w)$ scattering mode is associated to a creation operator. \par\noindent 
Now, being both the ``in" and ``out" basis complete, the ``in" and ``out" scattering modes are linearly related, namely 
\begin{eqnarray}\label{eq:outinaa}
  \phi_{vr}^{in}(w)&=& S_{ur,vr} \phi_{ur}^{out}(w)+S_{3l,vr} \phi_{3l}^{out}(w)+S_{4l,vr} \phi_{4l}^{out}(w)\ , \nonumber \\
  \phi_{ul}^{in}(w) &=& S_{ur,ul} \phi_{ur}^{out}(w)+S_{3l,ul} \phi_{3l}^{out}(w)+S_{4l,ul} \phi_{4l}^{out}(w)\ , \nonumber \\
  \phi_{vl}^{in} (w)&=& S_{ur,vl} \phi_{ur}^{out}(w)+S_{3l,vl} \phi_{3l}^{out}(w)+S_{4l,vl} \phi_{4l}^{out}(w)\ .
\end{eqnarray}
This leads to a nontrivial Bogoliubov transformation between the ``in" and ``out" creation and annihilation operators 
\begin{eqnarray}
\hat a_{ur}^{out}(w) &=& S_{ur,vr} \hat a_{vr}^{in}(w) + S_{ur,ul} \hat a_{ul}^{in}(w)+S_{ur,vl} \hat a_{vl}^{in\dagger}(w)\ , \nonumber \\
\hat a_{3l}^{out}(w) &=& S_{3l,vr} \hat a_{vr}^{in}(w) + S_{3l,ul} \hat a_{ul}^{in}(w)+S_{3l,vl} \hat a_{vl}^{in\dagger}(w)\ , \nonumber \\
\hat a_{4l}^{out\dagger}(w) &=& S_{4l,vr} \hat a_{vr}^{in}(w) + S_{4l,ul} \hat a_{ul}^{in}(w)+S_{4l,vl} \hat a_{vl}^{in\dagger}(w)\ . \label{smatcoef}
\end{eqnarray}
The $3\times 3$ scattering matrix $S$ whose elements are the $S_{ij,ml}$ coefficients of eqs. (\ref{smatcoef}) relate the ``in" and ``out" basis. One can show that it satisfies $S^\dagger \eta S=\eta = S\eta S^\dagger $ where the $3\times 3$ matrix $\eta$ is $\eta=diag(1,1,-1)$. \par\noindent
As clearly seen from eqs. (\ref{smatcoef}), the scattering matrix mixes creation and annihilation operators as a consequence of the fact that the basis contains both positive and negative frequency states and they get mixed passing from one basis to the other (see eqs. (\ref{eq:outinaa})). Because of this, the Hilbet space $H_{in}$, constructed out of the ``in" vacuum $|0,in\rangle$ (i.e. the state such that $\hat a_{ij}^{in}(w) |0,in\rangle =0$ for every $i,j$) by the action of creation $\hat a_{ij}^{in\dagger}(w)$ operators, and the 
corresponding $H_{out}$ Hilbert are not unitarily related. In particular $| 0,in\rangle\neq |0,out\rangle$. This means that, if we prepare the system to be at $t=-\infty$ in the ``in" vacuum state, $|0,in\rangle$, i.e. the state containing no incoming quanta, at late time ($t=+\infty$) this state (remember we work in the Heisenberg picture) will in general contain outgoing quanta. \par\noindent
The physical effect we are therefore describing is the spontaneous emission of Bogoliubov (quasi) particles out of the vacuum that shows up in our BEC with a white hole like flow. This proceeds by the conversion of vacuum fluctuations into real on shell particles. 
The number of the different outgoing particles created out of the vacuum can be computed using eqs.
(\ref {smatcoef}) as
\begin{eqnarray}
n_w^{ur}&=&\langle 0,in|\hat a_{ur}^{out\dagger}(w)\hat a_{ur}^{out}(w)|0,in\rangle =|S_{ur,vl}|^2\ ,\nonumber \\
n_w^{3l}&=&\langle 0,in|\hat a_{3l}^{out\dagger}(w)\hat a_{3l}^{out}(w)|0,in\rangle=|S_{3l,vl}|^2\ , \nonumber \\
n_w^{4l}&=&\langle 0,in|\hat a_{4l}^{out\dagger}(w)\hat a_{4l}^{out}(w)|0,in\rangle=|S_{4l,vr}|^2+
|S_{4l,ul}|^2\ . \label{sund} \end{eqnarray}
 Here $n_w^{ur}$ is (for given $w$) the number of $k_u^r$ particles emitted in the right (subsonic) region and travelling towards $x=+\infty$, $n_w^{3l}$ and $n_w^{4l}$ are respectively the number of $k_3^l$ and $k_4^l$ particles emitted in the left (supersonic) region: they travel supersonically with $|v_g|>v_0$ against the flow towards $x=-\infty$. 
Note that the $k_4^l$ quanta correspond to negative energy particles, they are the so called ``partners".
From unitarity it follows
\be n_w^{ur}+n_w^{3l}=n_w^{4l}\ , \ee
i.e. the number of positive energy particles equals the number of the negative energy ones as required by energy conservation. It is the existence of these partners in the spectrum of the physical states , that allows particles creation to occur even in a stationary setting, like ours, without contradicting energy conservation.
\par \noindent From eqs. (\ref{svl} , \ref{sul}, \ref{sur}) we can infer the leading (in $w$) behaviour of these quantities
\begin{eqnarray}
n_w^{ur} &=& \frac{c_r}{c_l} \frac{(v_0-c_l)^2}{(v_0+c_r)^2} + O(w)\ ,\nonumber \\
n_w^{3l} &=& \frac{(c_r-v_0)(v_0^2-c_l^2)^{3/2}(c_r+c_l)}{2c_l(c_r+v_0)(c_r-c_l)}\frac{1}{w}+ const.\ ,\nonumber \\
n_w^{4l} &=& \frac{(c_r-v_0)(v_0^2-c_l^2)^{3/2}(c_r+c_l)}{2c_l(c_r+v_0)(c_r-c_l)}\frac{1}{w}+ const.\ .
\label{emiwh}\end{eqnarray}

\subsection{White hole versus black hole emission}

In this section we shall discuss in detail the results for the particles emission obtained in the previous section for our white hole configuration and compare them to the ones obtained for a black hole configuration.
The two are simply related: reversing the direction of the BEC flow, one simply moves from a white hole configuration to a BH one as shown in Figs. 16, 17.

\begin{figure}
\centering \includegraphics[angle=0, height=1.5in] {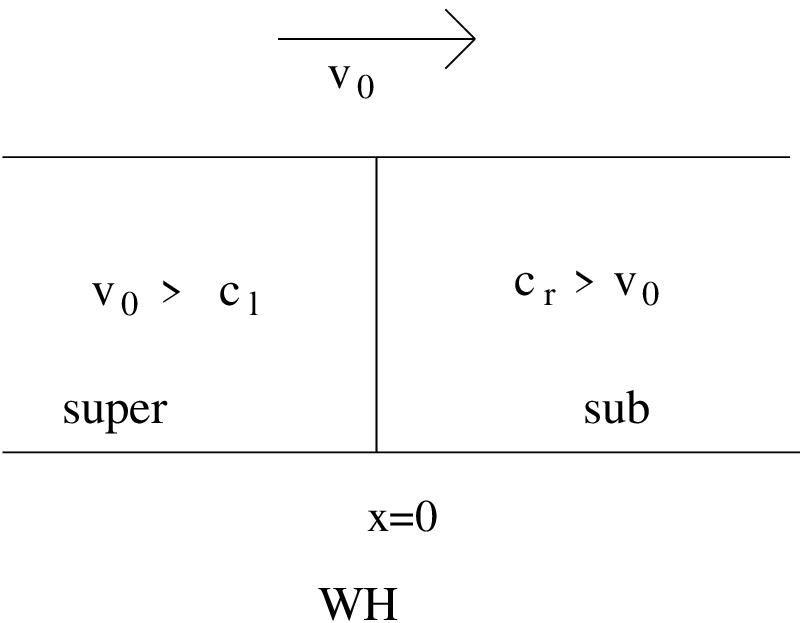}
\caption{White hole configuration.
}
\centering \includegraphics[angle=0, height=1.5in] {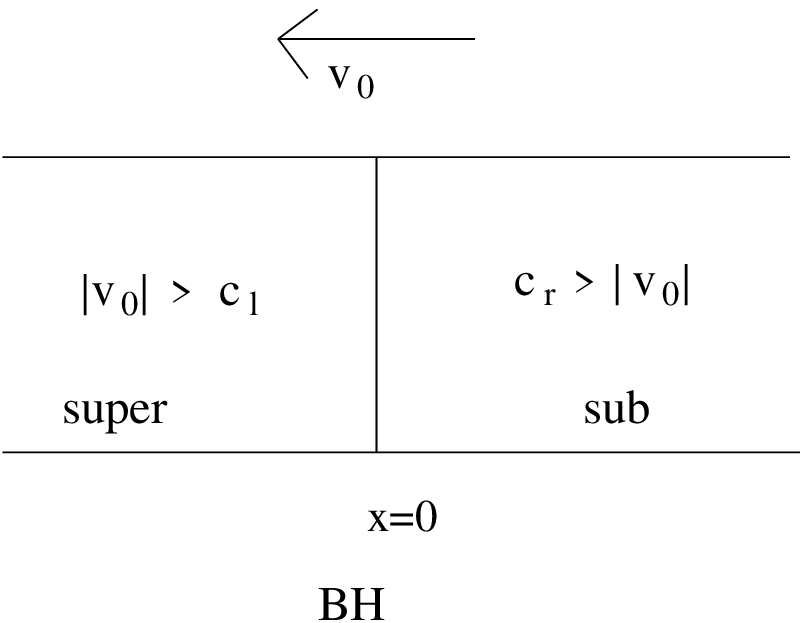}
\caption{Black hole configuration.
}
\end{figure}

For the black hole case sound waves are trapped in the $x<0$ supersonic region and dragged by the flow to the left, towards $x=-\infty$. The plot of the dispersion relation for the two configurations
\be w_0-v_0k=\pm\sqrt{c^2k^2+\frac{\xi^2k^4}{4}}\label{dddd}\ee
is represented in Figs. 18-21 .

\begin{figure}
\centering \includegraphics[angle=0, height=2.0in] {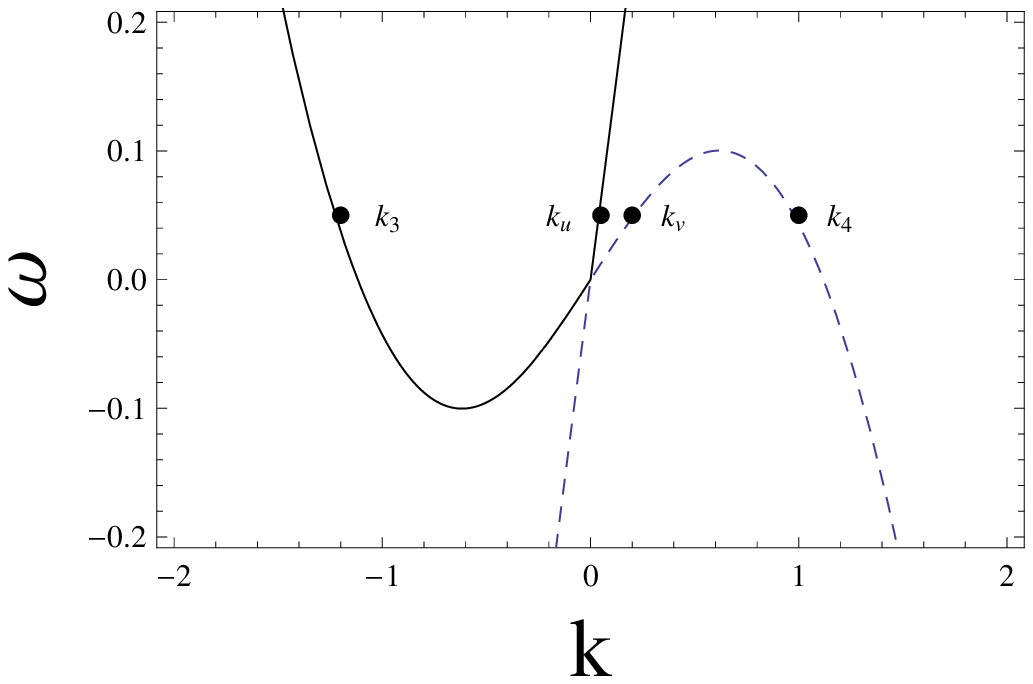}
\caption{Supersonic dispersion relation (left region) for the white hole.
}
\centering \includegraphics[angle=0, height=2.0in] {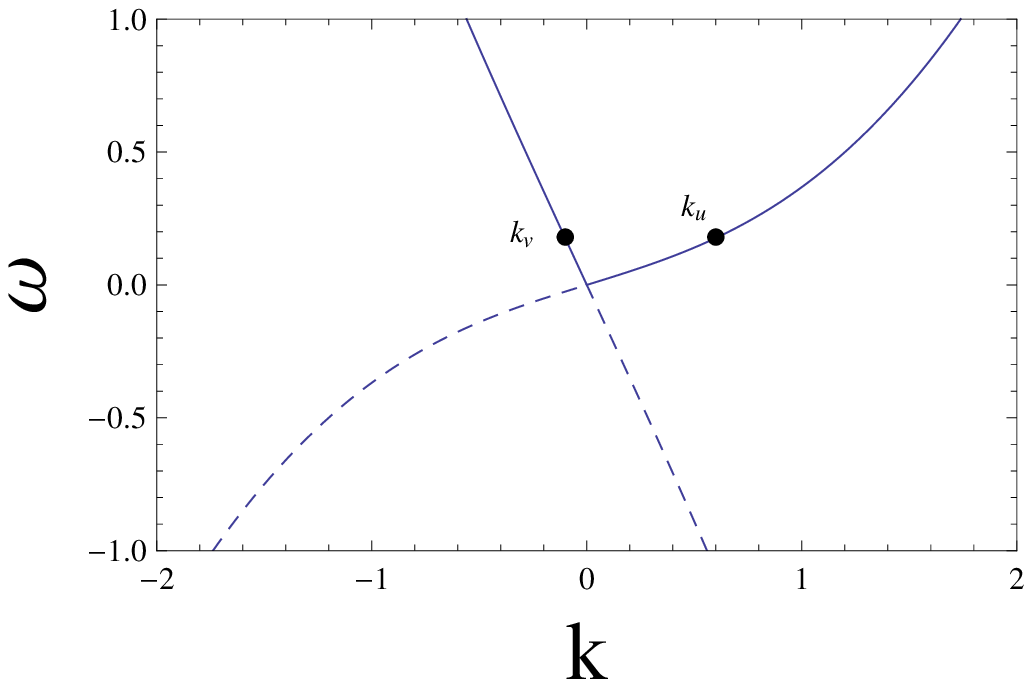}
\caption{Subsonic dispersion relation (right region) for the white hole.
}
\end{figure}

\begin{figure}
\centering \includegraphics[angle=0, height=2.0in] {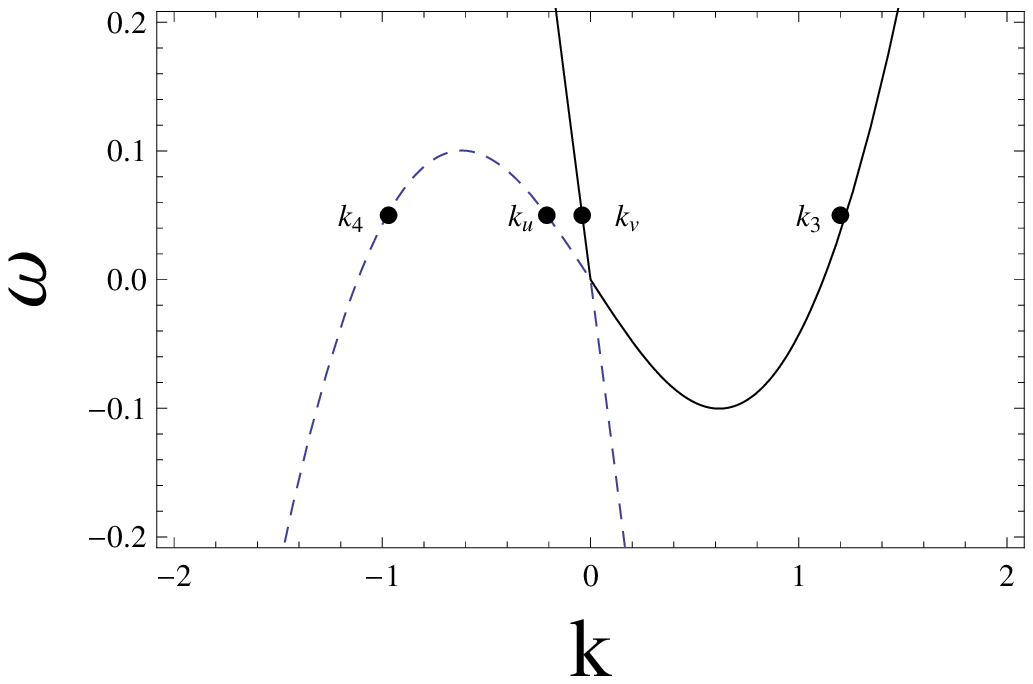}
\caption{Supersonic dispersion relation (left region) for the black hole.
}
\centering \includegraphics[angle=0, height=2.0in] {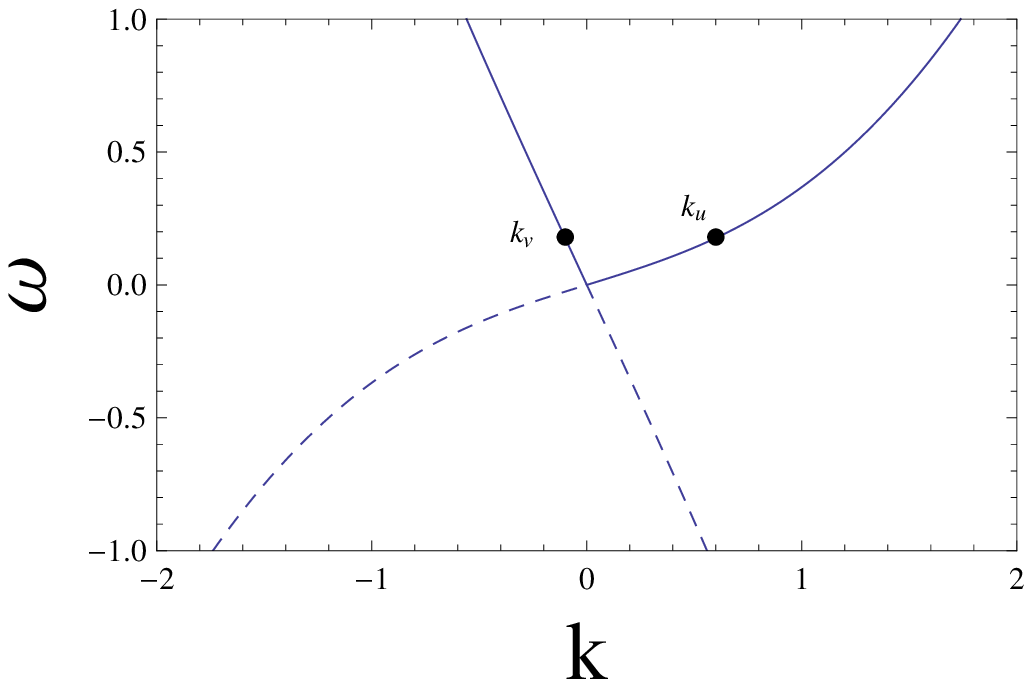}
\caption{Subsonic dispersion relation (right region) for the black hole.
}
\end{figure}

Moving from the WH configuration to the BH one ($v_0\to -v_0$) one sees from Figs. 22, 23  that the ingoing and outgoing character of the modes are exchanged, since BH and WH are related by time reversal and this operation reverses the sign of the group velocity.

\begin{figure}
\centering \includegraphics[angle=0, height=1.5in] {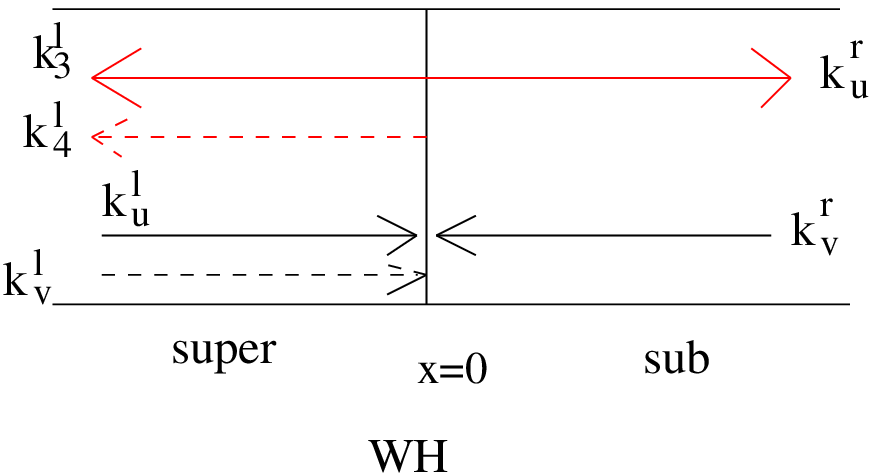}
\caption{'in' and 'out' modes in a white hole.
}
\centering \includegraphics[angle=0, height=1.5in] {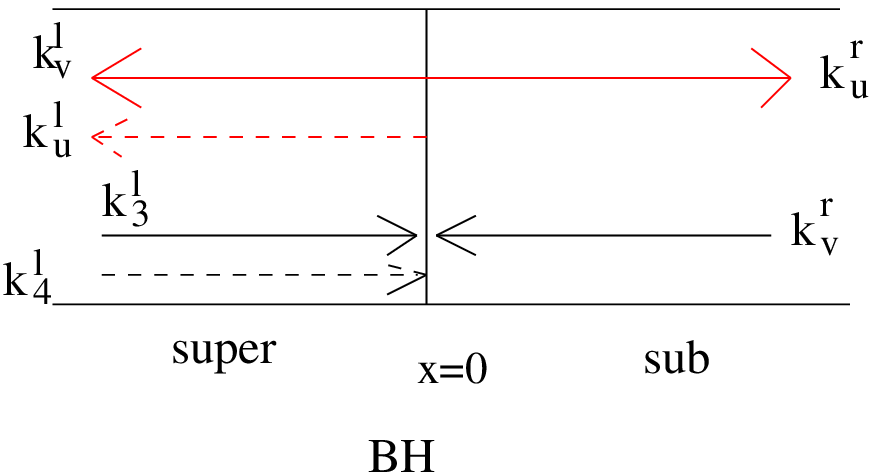}
\caption{'in' and 'out' modes in a black hole.
}
\end{figure}

They are connected by a $k\to -k$ symmetry as expected, since the dispersion relation eq. (\ref{dddd}) is invariant under the combined $k\to -k\ , v_0\to -v_0$ reflection. \par \noindent
Looking at the expression for the roots $k_i(w)$ for the BH case, given in the Appendix, and performing the above symmetry operation, one has the following correspondence
\begin{eqnarray} k_u^{r(l)}|_{WH} &\longleftrightarrow& k_v^{r(l)}|_{BH}\ , \nonumber \\
k_v^{r(l)}|_{WH} &\longleftrightarrow& k_u^{r(l)}|_{BH}\ , \nonumber \\
k_3^{l}|_{WH} &\longleftrightarrow& k_3^{l}|_{BH}\ , \nonumber \\
k_4^{l}|_{WH} &\longleftrightarrow& k_4^{l}|_{BH}\ . \label{corresp}
\end{eqnarray}
At the fundamental level of the solution of the basic dynamical equations (\ref{cde}) one can establish the correspondence between a WH and a BH mode as following \cite{macher}
\be \phi_j^{BH}(t,x) \to \phi_i^{WH}(t,x)=[\phi^{BH}_j(-t,x)]^*\ee
since under time reversal ($t\to -t$, $v_0\to -v_0$) eqs. (\ref{cde}) get mapped in their complex conjugated. The correspondence between the indices $i$ and $j$ is the one shown in (\ref{corresp}). 
Hence the WH ``in" scattering modes ${}^{WH}\phi^{in}_i(w)$ constituting the ``in" basis for the field operator expansion can be related to the ``out" BH scattering modes ${}^{BH}\phi_j^{out}(w)$ (i.e. the ``out" basis for the BH case) as follows
\begin{eqnarray}
{}^{WH}\phi_{vr}^{in}(t,x,v_0^{WH}) &=& \left[{}^{BH}\phi_{ur}^{out}(-t,x,v_0^{BH})\right]^*\ ,\nonumber \\
{}^{WH}\phi_{ul}^{in}(t,x,v_0^{WH}) &=& \left[{}^{BH}\phi_{vl}^{out}(-t,x,v_0^{BH})\right]^*\ ,\nonumber \\
{}^{WH}\phi_{vl}^{in}(t,x,v_0^{WH}) &=& \left[{}^{BH}\phi_{ul}^{out}(-t,x,v_0^{BH})\right]^*\ ,
\end{eqnarray}
where $v_0^{BH}=-v_0^{WH}$.
This implies the simple relation between the WH scattering matrix $S_{WH}$ and the corresponding BH one $S_{BH}$ 
\be \left[ S_{WH}\right]^* = \left[ S_{BH}\right]^{-1} \ , \ee
which can re rewritten as
\be S_{WH}=\eta \left[ S_{BH} \right]^T \eta \label{zz}\ee
where use of the unitarity of the scattering matrix ($S^T\eta S=\eta$) has been made. \par\noindent
Having established the relation between the $S$ matrices for WH and BH flow, we can now start to discuss our WH results of the previous section. As we have seen (eqs. (\ref{sund})) we have production of every three kind of particles. 
We begin with the emission into the (right) subsonic region, which can be considered as the exterior of the WH. Here we have emission of ($u,r$) quanta according to the first of eqs. (\ref{emiwh})
\be \label{whe} {}^{WH}n_w^{ur} = |S_{ur,vl}|^2 = \frac{c_r}{c_l}\frac{(v_0^{WH}-c_l)^2}{(v_0^{WH}+c_r)^2} + ..
\ee
where we have added a superscript ``$WH$" to emphasyze that the flow corresponds to a WH. The relevant channel responsible for this emission is in Fig. 24. The associated channel for a BH, according to the correspondence of (\ref{corresp}) is given in  Fig. 25 and has an amplitude (see Appendix)
\be S^{BH}_{ul,vr}=\sqrt{\frac{c_r}{c_l}} \frac{v_0^{BH}+c_l}{c_r-v_0^{BH}}=\sqrt{\frac{c_r}{c_l}}\frac{c_l-v_0^{WH}}{c_r+v_0^{WH}}=-S_{ur,vl}^{WH}\label{df}\ee
in accordance with (\ref{zz}). \par\noindent

\begin{figure}
\centering \includegraphics[angle=0, height=1.5in] {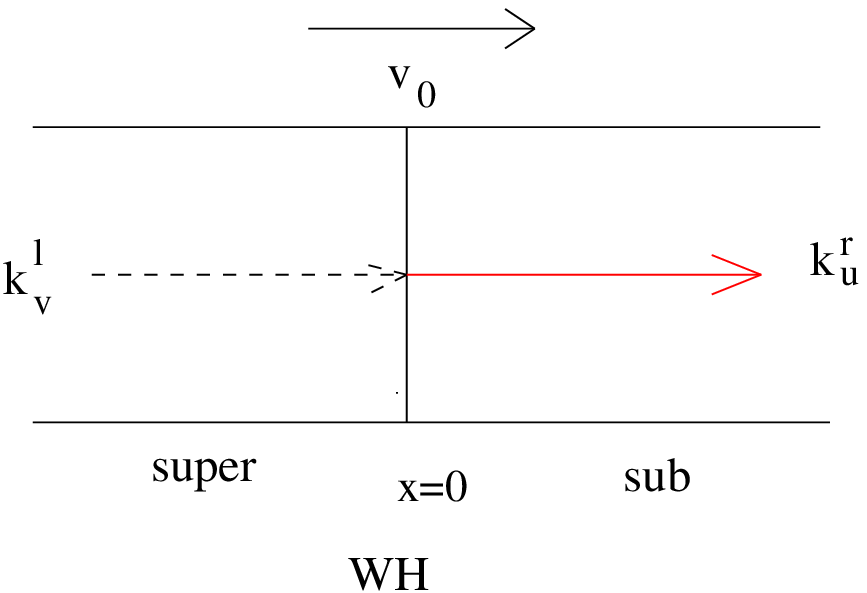}
\caption{Relevant channel for the emission in the white hole exterior.
}
\centering \includegraphics[angle=0, height=1.5in] {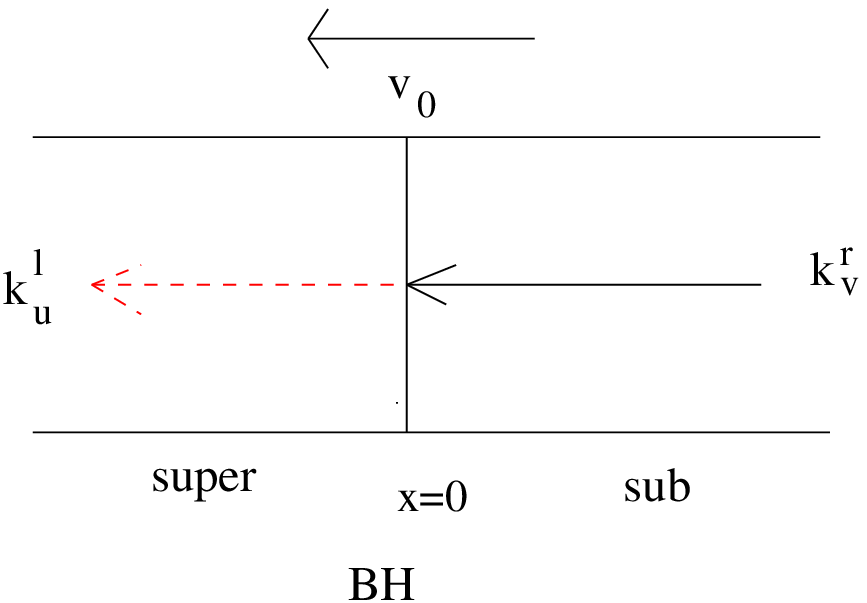}
\caption{Associated black hole channel to that in Fig. 24.
}
\end{figure}

Now eq. (\ref{whe}) reveals that the emission from a WH in the subsonic region has at small $w$ a flat spectrum. It has not the $1/w$ thermal behaviour that characterizes the Hawking like emission from a BH flow. 
As seen from eq. (\ref{df}) this $WH$ $k_u^r$ emission is related to the emission of negative energy $k_u^l$ ``partners" inside a BH. 

Hawking radiation from a BH is described by the channel in Fig. 26 and the corresponding S-matrix element 
(see Appendix) gives
\be {}^{BH}n_w^{ur}=|S_{ur,4l}^{BH}|^2\simeq  \frac{(v_0^{BH 2}-c_l^2)^{3/2}(v_0^{BH}+c_r)}{(c_r^2-c_l^2)(c_r-v_0^{BH})}\frac{2c_r}{w} \ ,\ee
which describes a (approximately) thermal emission at an effective temperature 
 \be T=\frac{2c_r}{k_B}\frac{(v_0^{BH 2}-c_l^2)^{3/2}(v_0^{BH}+c_r)}{(c_r^2-c_l^2)(c_r-v_0^{BH})}\ .\ee

Being (see eq. (\ref{zz})) $S_{ur,4l}^{BH}=-S_{4l,vr}^{WH}$, one sees that in a WH the process associated to Hawking like thermal emission occurring in a BH is the production of $k_4^l$ negative energy ``partners"  
according to the channel in Fig 27.

\begin{figure}
\centering \includegraphics[angle=0, height=1.5in] {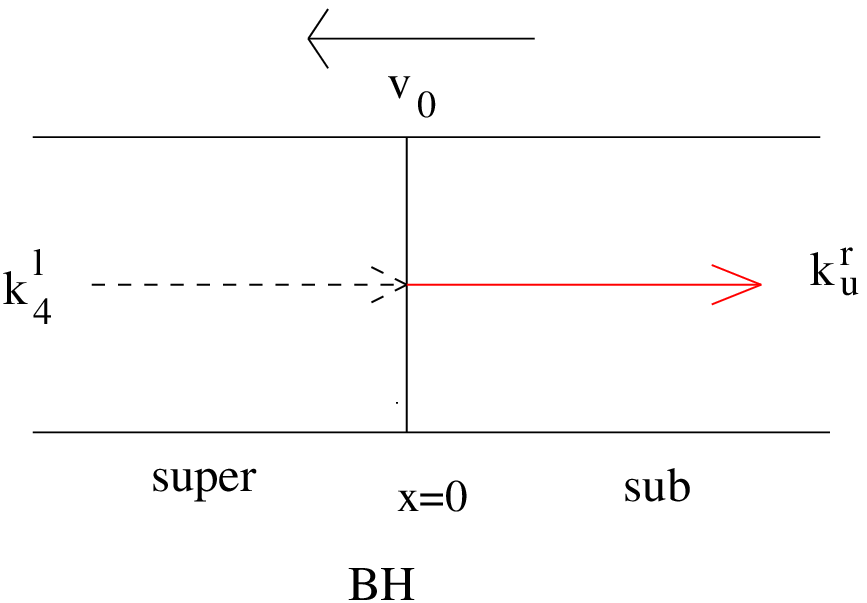}
\caption{Channel responsible for black hole emission.
}
\centering \includegraphics[angle=0, height=1.5in] {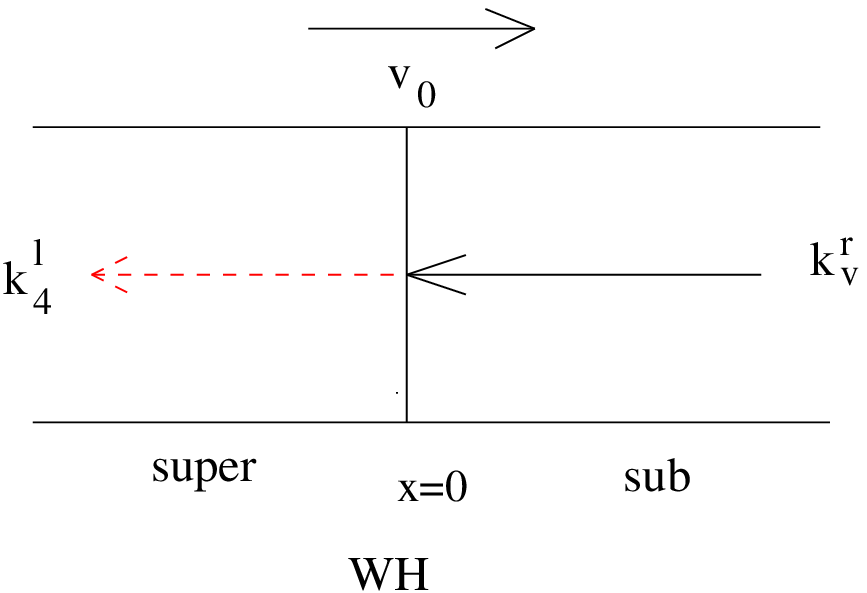}
\caption{White hole channel associated to that in Fig. 26.
}
\end{figure}

 Finally, the production of positive energy $k_3^l$ quanta inside a WH, Fig. 28, with $n_w^{3l}=|S_{3l,vl}^{WH}|^2$ (see the second of eqs. (\ref{sund}) and (\ref{emiwh})) 
is related to the following BH process Fig. 29, namely the production of $k_u^l$ partners inside the BH described by $S_{ul,3l}^{BH}$ with $S_{ul,3l}^{BH}=(-S_{3l,vl}^{WH})^*$ according to eq. (\ref{zz}). 

\begin{figure}
\centering \includegraphics[angle=0, height=1.5in] {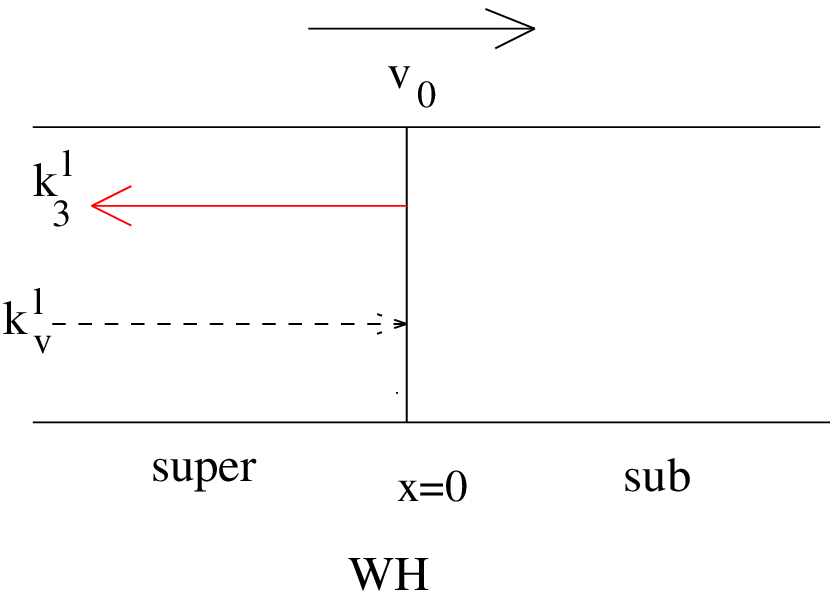}
\caption{Channel associate to the production of positive energy $k_3^l$ quanta inside a 
white hole.
}
\centering \includegraphics[angle=0, height=1.5in] {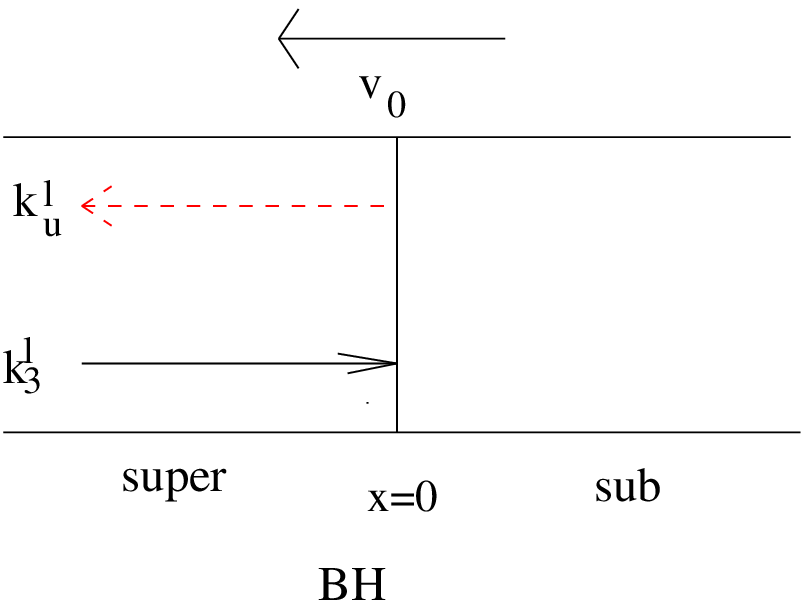}
\caption{Black hole channel associated to that in Fig. 28. 
}
\end{figure}

\section*{Conclusions}

We have seen that although the scattering matrices for BECs BHs and WHs quasi particle production are simply algebraically related ($S_{WH}^*=S_{BH}^{-1}$), the physical features of the corresponding emission are rather different in the two contexts.

The emission by a BEC WH in the subsonic region (i.e. outside the `antitrapping' region bounded by the sonic horizon) has a flat spectrum that contrasts the thermal character of the BH emission in the subsonic region, which is the analogue of the process discovered by Hawking in the gravitational setting.
A similar process indeed exists for BEC WHs but occurs inside the horizon. The close relation existing between the two processes is summarized in $S^{BH}_{ur,4l}=-S^{WH}_{4l,vr}$.
However the produced quasiparticles are characterized by quite different modes in the two cases: a hydrodynamical `u' mode in the BH case and a dispersive `$k_4$' mode in the WH one.

This leads to striking different physical features of the corresponding emissions. In particular the associated density correlations functions inside WHs show a peculiar growing checkerboard pattern \cite{wh} which has no analogue in the BH case (unless one considers a nonvanishing momentum transverse to the flow \cite{cfpba}). 
The responsible of this instability is a zero frequency propagating outgoing mode which is absent 
in the BH radiation ($\lim_{w\to 0} k_u^{r\ BH}=0,\ \lim_{w\to 0} k_4^{l\ WH}\neq 0$) as can be seen in Figs. 18, 20. 

What can we infer from our analysis with respect to gravitational WHs (if these objects indeed exist)? 
First one should say that particle production by gravitational WHs has not been a subject much investigated leading however to controversial results. 
Just after Hawking's discovery in '74 of BH radiation Wald \cite{wald} analyzed the WH context concluding that the emission becomes singular. On the other hand, Hawking in \cite{Hawking:1976de} claims
that BHs and WHs behave exactly in the same way, i.e. also WHs should emit thermal radiation. Other claims that WHs `antievaporate' are also present in the literature (see for instance \cite{schutzhold}). One should say that the results, unlike the BH case, seem to depend heavily on the WH formation process. Not to say about the role played by the past singularity and the boundary condition on it.

Unfortunately our work like others on BEC WHs pseudo particles radiation does not shed light on the gravitational counterpart . The reason for this is that, as we have seen, the thermal like emission of BEC WHs is associated to the modes ($k_3^{l\ WH},\ k_4^{l\ WH}$) which are highly dispersive. The associated quanta differ significantly from relativistic massless quanta which are characterized by a non dispersive linear relation between frequency $w$ and wavevector $k$. The $k_{3,4}$ modes are not hydrodynamical modes. The situation is just the opposite of what happens in BEC BHs thermal emission whose modes $k_u^{r\ BH},\ k_u^{l\ BH}$ are hydrodynamical and obey linear dispersion relation.

Hence WH BECs thermal emission is, unlike the BH case, highly sensible to the dispersion. One can say that it feels the short distance features of the underlying quantum theory. In this sense the 
robustness of BH radiation \cite{finpar} does not extend to WHs.

One should always remember that the so called analogy between gravity and some condensed matter systems relies on the possibility of treating the condensed matter system under the hydrodynamical approximation where modes are non dispersive and obey a linear `relativistic' dispersion relation. 
In view of all this one should hardly consider BEC WHs pseudoparticle production as a reliable insight for the corresponding  gravitational setting. 

\acknowledgments

We thank I. Carusotto for many useful and interesting discussions.

\newpage

\appendix

\section{Black hole S-matrix coefficients}

In this appendix we recall the $k_i(w)$ modes for the black hole, in the right ($r$) subsonic and left ($l$) supersonic regions
\begin{eqnarray}\label{uv}
&&k_v^r=\frac{\omega}{v_0-c_r}\left(1+\frac{c_rw^2\xi_r}{8(v_0-c_r)^3}+O(z_r^4)\right)\ ,\nonumber\\
&&k_u^r=\frac{\omega}{v_0+c_r}\left(1-\frac{c_rw^2\xi_r}{8(v_0+c_r)^3}+O(z_r^4)\right)\ ,
\nonumber \\ 
&&k_v^l=\frac{\omega}{v_0-c_l}\left[1+\frac{c_lw^2\xi_l^2}{8(v_0-c_l)^3}+O(z_l^2)\right]\ ,\nonumber\\
&&k_u^l=\frac{\omega}{v_0+c_l}\left[1-\frac{c_lw^2\xi_l^2}{8(v_0+c_l)^3}+O(z_l^2)\right]\ ,
\nonumber \\ 
&& k_{3(4)}^l=\frac{\omega v_0}{c_l^2-v_0^2}\left[1-\frac{(c_l^2+v_0^2)c_l^2w^2\xi_l^2}{4(c_l^2-v_0^2)^3}+O(z_l^4)\right]
+(-)\frac{2\sqrt{v_0^2-c_l^2}}{c_l\xi_l}\left[1+\frac{(c_l^2+2v_0^2)c_l^2w^2\xi_l^2}{8(c_l^2-v_0^2)^3}+O(z_l^4)\right]\ \  \nonumber 
\end{eqnarray}
and the S-marix coefficients: 
\begin{eqnarray}\label{3in}
 S_{vl,vr}^{BH}  &=& \sqrt{\frac{c_r}{c_l}}\frac{v_0-c_l}{v_0-c_r} \ , \nonumber \\
 S_{ur,vr} ^{BH} &=& \frac{v_0+c_r}{v_0-c_r} \ ,
\nonumber\\
  S_{ul,vr}^{BH} &=&  \sqrt{\frac{c_r}{c_l}}\frac{v_0+c_l}{c_r-v_0}  \ ,
\end{eqnarray}
for the left moving `in' mode from the subsonic region in Fig. 30;

\begin{eqnarray}\label{3in}
 S_{vl,3l}^{BH} &=& \frac{(v_0^2-c_l^2)^{3/4}(v_0+c_r)}{\sqrt{2c_lw}(c_l+c_r)\sqrt{c_r^2-v_0^2}}\left(\sqrt{c_r^2-v_0^2}+i\sqrt{v_0^2-c_l^2}\right)
\ \ ,
\nonumber \\
 S_{ur,3l} ^{BH} &=& \frac{\sqrt{2c_r}(v_0^2-c_l^2)^{3/4}(v_0+c_r)}{\sqrt{w}(c_r^2-c_l^2)\sqrt{c_r^2-v_0^2}}\left(\sqrt{c_r^2-v_0^2}+i\sqrt{v_0^2-c_l^2}\right)
\ \ , \nonumber\\
S_{ul,3l}^{BH}  &=& \frac{(v_0^2-c_l^2)^{3/4}(v_0+c_r)}{\sqrt{2c_lw}(c_l-c_r)\sqrt{c_r^2-v_0^2}}
\left(\sqrt{c_r^2-v_0^2}+i\sqrt{v_0^2-c_l^2}\right)\ \ 
\end{eqnarray}

for the positive norm right moving `in' mode,  Fig. 31, and finally

\begin{eqnarray}\label{4in}
  S_{vl,4l}^{BH} &=& \frac{(v_0^2-c_l^2)^{3/4}(v_0+c_r)}{\sqrt{2c_lw}(c_l+c_r)\sqrt{c_r^2-v_0^2}}\left(\sqrt{c_r^2-v_0^2}-i\sqrt{v_0^2-c_l^2}\right)\ \ \ , \\
  S_{ur,4l}^{BH} &=& \frac{\sqrt{2c_r}(v_0^2-c_l^2)^{3/4}(v_0+c_r)}{\sqrt{w}(c_r^2-c_l^2)\sqrt{c_r^2-v_0^2}}\left(\sqrt{c_r^2-v_0^2}-i\sqrt{v_0^2-c_l^2}\right) \ \ ,\\
 S_{ul,4l}^{BH}  &=& \frac{(v_0^2-c_l^2)^{3/4}(v_0+c_r)}{\sqrt{2c_lw}(c_l-c_r)\sqrt{c_r^2-v_0^2}}\left(\sqrt{c_r^2-v_0^2}-i\sqrt{v_0^2-c_l^2}\right)\ \  \ 
\end{eqnarray}

for the negative norm `in' mode from the left supersonic region.
\begin{figure}
\centering \includegraphics[angle=0, height=1.5in] {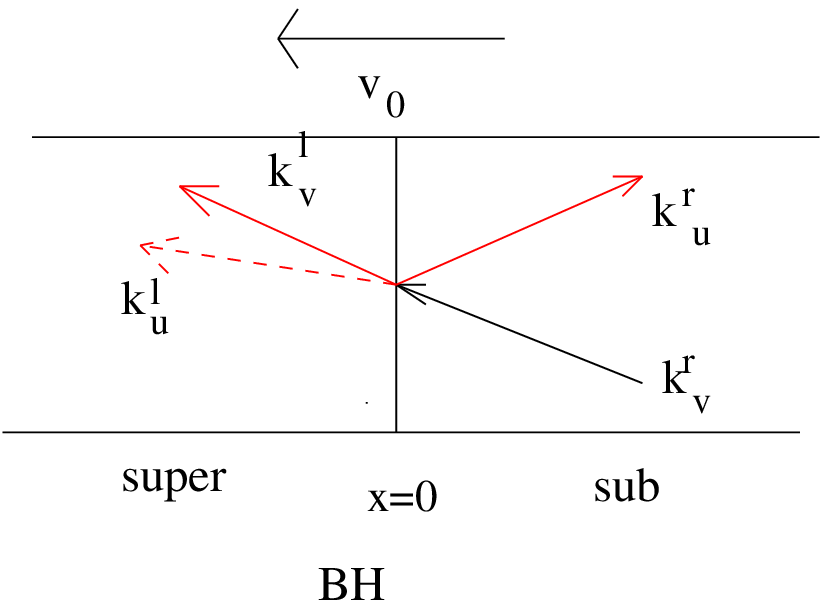}
\caption{Incoming mode from the right subsonic region. }
\centering \includegraphics[angle=0, height=1.5in] {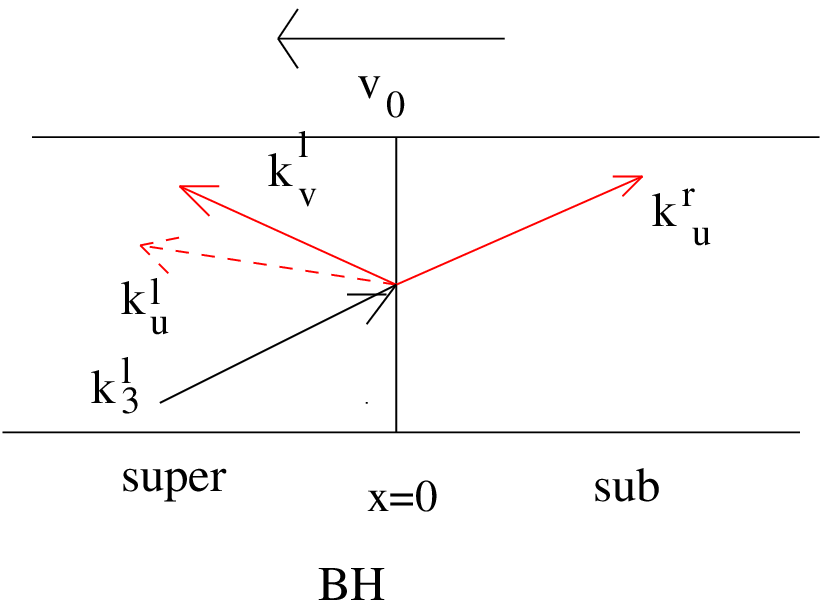}
\caption{Positive norm `in' mode from the left supersonic region. }
\centering \includegraphics[angle=0, height=1.5in] {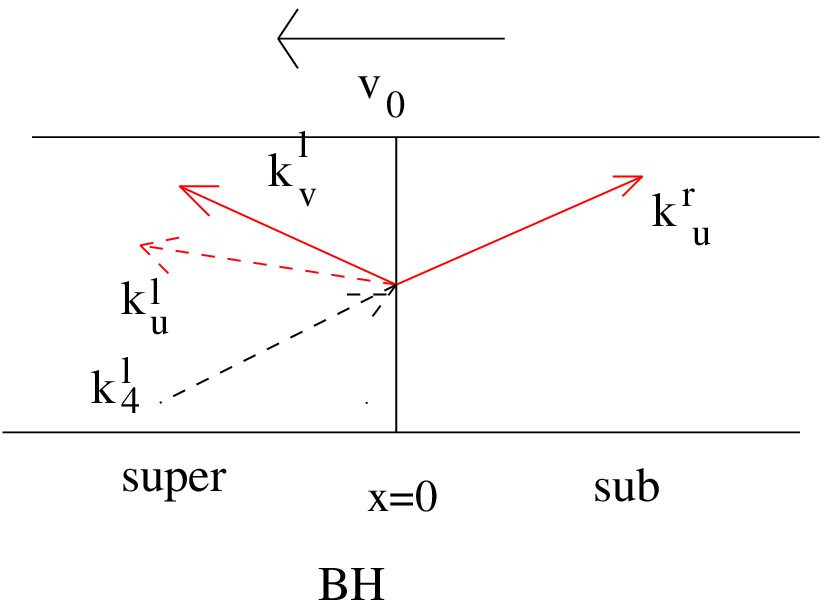}
\caption{Negative norm `in' mode from the left supersonic region. }
\end{figure}


\begin{thebibliography}{99}
\bibitem{hawking}
S. Hawking, {\it Nature} {\bf 248} (1974), 30; {\it Commun. Math. Phys.} {\bf 43} (1975), 199
\bibitem{mapar}
S. Massar and R. Parentani, {\it Phys. Rev.} {\bf D54} (1996), 7444
\bibitem{jacobson}
T. Jacobson, {\it Phys. Rev.} {\bf D44} (1991), 1731
\bibitem{eardley}
D.M. Eardley, {\it Phys. Rev. Lett.} {\bf 33} (1974), 442
\bibitem{unruh}
W. Unruh, Phys. Rev. Lett. {\bf 46}, 1351 (1981)
\bibitem{Hawking:1973uf} 
  S.~W.~Hawking and G.~F.~R.~Ellis,
  ``The Large scale structure of space-time,''
  Cambridge University Press, Cambridge, 1973
\bibitem{pistri}
L. Pitaevski and S. Stringari, {\it Bose Einstein condensation} (Oxford University Press, 2003) 
\bibitem{wh}
C. Mayoral, A. Recati, A. Fabbri, R. Parentani, R. Balbinot and I. Carusotto, {\it New J. Phys.} {\bf 13} (2011), 025007
\bibitem{step}
A. Recati, N. Pavloff and I. Carusotto, Phys. Rev. {\bf A80}, 043603 (2009);  C. Mayoral, A. Fabbri and M. Rinaldi, Phys. Rev. {\bf D83}, 124047 (2011); P.E. Larr\'e, A. Recati, I. Carusotto and N. Pavloff, {\it Phys. Rev.} {\bf A85} (2012), 013621
\bibitem{Balbinot:2012xw} 
  R.~Balbinot, I.~Carusotto, A.~Fabbri, C.~Mayoral and A.~Recati,
  ``Understanding Hawking radiation from simple models of atomic Bose-Einstein condensates,''
  arXiv:1207.2660 [gr-qc].
\bibitem{macher}
J. Macher and R. Parentani, {\it Phys. Rev.} {\bf D79} (2009), 124008
\bibitem{cfpba}
A. Coutant, A. Fabbri, R. Parentani, R. Balbinot and P. Anderson, {\it Phys. Rev.} {\bf D86} (2012), 064022
\bibitem{wald}
R.M.  Wald ad S. Ramaswany, {\it Phys. Rev.} {\bf D21} (1980), 2736
\bibitem{Hawking:1976de} 
  S.~W.~Hawking,
  Phys.\ Rev.\ D {\bf 13}, 191 (1976).
\bibitem{schutzhold}
R. Schutzhold, {\it Phys. Rev.} {\bf D64} (2001), 024029
\bibitem{finpar}
S. Finazzi and R. Parentani, {\it Phys. Rev.} {\bf D83} (2011), 084010

\end{thebibliography}
\end{document}